\documentclass[sigconf]{acmart}
\usepackage{booktabs} 
\usepackage{float}
\usepackage{multirow}
\usepackage{graphicx}
\usepackage{array}
\usepackage{xcolor}
\usepackage{soul}
\usepackage{wrapfig}
\usepackage[utf8]{inputenc}
\usepackage{subcaption}
\usepackage{xfrac}
\usepackage{listings}
\usepackage{fancybox}

\newcommand{\ignore}[1]{}
\newcolumntype{P}[1]{>{\centering\arraybackslash}p{#1}}
\newcolumntype{M}[1]{>{\centering\arraybackslash}m{#1}}
\AtBeginDocument{%
  }

\copyrightyear{2026}
\acmYear{2026}
\setcopyright{cc}
\setcctype{by-nc-nd}
\acmConference[CHI '26]{Proceedings of the 2026 CHI Conference on Human Factors in Computing Systems}{April 13--17, 2026}{Barcelona, Spain}
\acmBooktitle{Proceedings of the 2026 CHI Conference on Human Factors in Computing Systems (CHI '26), April 13--17, 2026, Barcelona, Spain}
\acmPrice{}
\acmDOI{10.1145/3772318.3791495}
\acmISBN{979-8-4007-2278-3/2026/04}

\acmISBN{978-1-4503-XXXX-X/18/06}

\begin{document}
\setlength{\tabcolsep}{2pt}

\title{Creating Disability Story Videos with Generative AI: Motivation, Expression, and Sharing}


\author{Shuo Niu}
\email{shniu@clarku.edu}
\orcid{0000-0002-8316-4785}
\affiliation{%
  \institution{Clark University}
  \streetaddress{950 Main Street}
  \city{Worcester}
  \state{Massachusetts}
  \country{USA}
  \postcode{01610}
}

\author{Dylan Clements}
\email{dyclements@clarku.edu}
\orcid{0009-0003-4672-1524}
\affiliation{%
  \institution{Clark University}
  \streetaddress{950 Main Street}
  \city{Worcester}
  \state{Massachusetts}
  \country{USA}
  \postcode{01610}
}

\author{Hyungsin Kim}
\email{HyuKim@clarku.edu}
\orcid{0000-0002-3794-1686}
\affiliation{%
  \institution{Clark University}
  \streetaddress{950 Main Street}
  \city{Worcester}
  \state{Massachusetts}
  \country{USA}
  \postcode{01610}
}
\renewcommand{\shortauthors}{Niu et al.}

\begin{abstract}
  Generative AI (GenAI) is both promising and challenging in supporting people with disabilities (PwDs) in creating stories about disability. GenAI can reduce barriers to media production and inspire the creativity of PwDs, but it may also introduce biases and imperfections that hinder its adoption for personal expression. In this research, we examine how nine PwD from a disability advocacy group used GenAI to create videos sharing their disability experiences. Grounded in digital storytelling theory, we explore the motivations, expression, and sharing of PwD-created GenAI story videos. We conclude with a framework of momentous depiction, which highlights four core affordances of GenAI that either facilitate or require improvements to better support disability storytelling: non-capturable depiction, identity concealment and representation, contextual realism and consistency, and emotional articulation. Based on this framework, we further discuss design implications for GenAI in relation to story completion, media formats, and corrective mechanisms.
\end{abstract}



\received{20 February 2007}
\received[revised]{12 March 2009}
\received[accepted]{5 June 2009}

\maketitle

\section{Introduction}
Digital storytelling with short-form videos has been a critical pathway for many people with disabilities (PwDs) to share their experiences and advocate for disability rights~\cite{gleason_future_2020, saridaki_digital_2018}. Extensive HCI and accessibility research shows that creating and sharing disability-related videos enables PwDs to present authentic portrayals of their lives~\cite{hibbard_vlogging_2011, seo_understanding_2018, borgos-rodriguez_myautsomefamilylife_2019}, express creativity and personal narratives~\cite{duval_chasing_2021, li_it_2022}, share disability-related challenges and solutions~\cite{niu_please_2024}, and build supportive communities~\cite{huh_health_2014, lyu_i_2024}. Storytelling practices allow PwDs to reflect on their identities~\cite{lyu_i_2024} and imagine accessible and just futures~\cite{angelini_criptopias_2023}. Sharing stories on platforms such as YouTube and TikTok also enables PwDs to challenge ableist assumptions~\cite{rodger_journeycam_2019, hofmann_living_2020} and engage with non-disabled audiences~\cite{niu_please_2024, choi_its_2022, duval_chasing_2021}. However, creating story videos requires decisions about storylines, the expression of emotions and moments, the crafting of visual and narrative media, the assembly of these components, and the sharing of the final video with others~\cite{lambert_digital_2013}. Video-based disability storytelling is often hindered by accessibility barriers in learning media production tools, communication challenges, and ableist perspectives from audiences~\cite{lyu_i_2024, niu_please_2024, choi_its_2022, heung_vulnerable_2024}.

\par
To support disability storytelling, HCI research has recently explored the integration of Generative Artificial Intelligence (GenAI), including large language models (LLMs) and AI-based visual and audio generators, as a means of enhancing creative expression~\cite{bennett_painting_2024, choi_exploring_2025, lee_altcanvas_2024}. However, their acceptance by PwDs and the affordances for creating diverse forms of story presentation remain underexplored. On the one hand, GenAI has been recognized as a valuable resource for learning, self-care, and self-advocacy~\cite{atcheson_id_2025}. GenAI has the potential to reduce media production barriers~\cite{bennett_painting_2024, saha_tutoria11y_2023} and foster independent content creation~\cite{choi_exploring_2025, lee_altcanvas_2024, mcnally_disability_2024}. On the other hand, researchers emphasize that GenAI often misrepresents PwDs and that its biases against them need to be addressed~\cite{glazko_autoethnographic_2023, mack_they_2024}. Disability-related content often perpetuates ableist stereotypes~\cite{s_guedes_artistic_2024} or reinforces ``inspirational'' narratives~\cite{bennett_painting_2024, gadiraju_i_2023, gogolushko_ai_2022}. 
\par

Disability storytelling videos can serve as respectful and impactful forms of advocacy~\cite{trevisan_crowd-sourced_2017, huh_health_2014, niu_please_2024}. However, such disclosure often involves emotional distress due to hate or harassment on social media~\cite{sannon_disability_2023, heung_vulnerable_2024}. Designs must therefore consider whether and how technologies support disability disclosure~\cite{porter_filtered_2017, zhang_its_2022}. GenAI offers an emerging approach for enabling personal expression and disability identity presentation in video-based storytelling while mitigating challenges tied to public visibility. However, it remains unclear whether GenAI can effectively support PwDs who lack experience in media technologies but wish to engage in disability activism and create online videos. HCI theories on GenAI emphasize designing for variability, co-creation, and imperfection~\cite{weisz_design_2024}, yet it is still unknown how GenAI's variability shapes PwDs' creation of disability stories, how they co-construct narratives with GenAI, and whether they choose to share the AI-generated content.

\par
In this seminal study, we present our experiences collaborating with members of a disability advocacy group who had limited prior experience with GenAI but were introduced to GenAI tools to create disability storytelling videos. Unlike prior studies that primarily focused on PwDs' interactions with GenAI chatbots and the evaluation of output quality~\cite{mack_they_2024, glazko_autoethnographic_2023}, we situate GenAI usage within their lived experiences of video-making in digital storytelling. This approach enabled PwDs to actively lead the co-creation process. It also allowed us to examine the core affordances of GenAI and identify potential challenges in supporting the articulation of everyday personal narratives. Moreover, because video-making requires assembling multiple GenAI outputs, we examined unique factors influencing the overall quality of the resulting stories. Drawing on Lambert's theory of digital storytelling~\cite{lambert_digital_2013}, we investigate the types of stories and moments PwDs are motivated to create with GenAI, the key factors shaping their expression through GenAI outputs, and their willingness and motivators to share GenAI-generated videos. Specifically, we address three research questions:

\begin{itemize}
    \item[RQ1] What disability-related moments and motivations do PwDs seek to depict when creating stories with GenAI?
    \item[RQ2] What components of stories do PwDs co-create with GenAI to express their disability experiences?
    \item[RQ3] How do PwDs decide whether to share their GenAI-created disability stories, and what motivates their sharing intentions?
\end{itemize}

In this research, we observed nine participants with disabilities from an advocacy group creating story videos about their lived experiences using GenAI. PwDs used ChatGPT to generate a six-scene video script, DALL-E to create AI images, and ElevenLabs to produce AI voiceovers for each scene. The images and voiceovers were then assembled into a final $\sim$1-minute video. Through qualitative analysis of GenAI prompts and semi-structured interviews, this study contributes a framework of \textit{momentous depiction} to guide the development of GenAI affordances for supporting disability storytelling. This framework highlights both the opportunities and challenges of GenAI in enhancing media efficacy. GenAI enables the portrayal of non-capturable moments and allows video production without disclosing one's identity. PwDs aimed to depict environments tied to their lived experiences where inaccessibility occurs and evoke emotional moments. However, personal stories could not be accurately conveyed by GenAI when misrepresentations occurred, such as inaccurate portrayals of disability or incorrect presentation of assistive technologies. The GenAI tools often produced generic or inconsistent characters and settings, thereby diminishing the authenticity.


\par

\section{Related Work}

\subsection{Storytelling in Disability Research}
\textit{Storytelling} in Human-Computer Interaction (HCI) research refers to users' reflection, expression, documentation, and reinterpretation of personal narratives and lived experiences~\cite{farao_digital_2023, taylor_cruising_2024, halperin_probing_2023, saksono_reflective_2017}. In HCI, storytelling practices and related technologies are studied to support PwDs in reflecting on disability narratives, communicating with others, and fostering imagination and learning~\cite{saha_tutoria11y_2023, nevsky_lights_2024, hofmann_living_2020, yoo_understanding_2021, yoo_remembering_2024}.
\par
Storytelling is also a reflective practice through which PwDs integrate lived experiences to express disability identities~\cite{hofmann_living_2020}. Prior HCI work shows that PwDs recall significant moments to convey emotions and share memories~\cite{yoo_understanding_2021, yoo_remembering_2024}. Students disclose disability-related narratives to access academic support~\cite{iniesto_designing_2020}. Blind individuals share meaningful moments with sighted loved ones through everyday storytelling~\cite{yoo_understanding_2021, yoo_remembering_2024} and foster family connections through creative expression~\cite{chheda-kothary_engaging_2024}. Deaf individuals use video blogging to communicate experiences~\cite{hibbard_vlogging_2011}. Individuals with mobility disabilities share wheelchair experiences to challenge inaccessible environments~\cite{rodger_journeycam_2019}. Neurodivergent individuals use social media to express personal interests and aspirations~\cite{bayor_leveraging_2019} and employ design-driven practices to reflect on their strengths and needs~\cite{boyd_exploring_2024}.

\par
Storytelling is a critical means of communication and raising disability awareness, but it also involves decisions about whether and how PwDs disclose their disabilities. On social media, storytelling enables PwDs to share emotions, build community~\cite{seo_understanding_2018, cocq_self-representations_2020}, and disclose personal challenges~\cite{niu_please_2024, seo_understanding_2018}. However, PwDs frequently encounter discrimination and exclusion online~\cite{miller_my_2017} and must navigate tensions between anonymity and authentic self-expression~\cite{mcclimens_presentation_2008, heung_vulnerable_2024}. Disclosure is often strategic and limited to trusted communities~\cite{furr_strategic_2016} as a way to avoid harassment~\cite{sannon_disability_2023, heung_vulnerable_2024}. Individuals with non-apparent disabilities may also face skepticism~\cite{bitman_which_2023}. Moreover, algorithmic curation can further marginalize disabled creators by restricting the visibility of their content~\cite{choi_its_2022, sannon_disability_2023, rauchberg_shadowbanned_2022}. Therefore, in online and virtual social contexts, PwDs strategically choose virtual personas (\textit{``selective disclosure''})~\cite{zhang_its_2022, porter_filtered_2017} and adopt varied self-presentation strategies~\cite{gualano_i_2024} to disclose or conceal their real-world disability conditions.
\par

HCI research also demonstrates that storytelling empowers PwDs by imagining possible futures and supporting skill development. Fictional and imaginative storytelling enables PwDs to envision accessible and just futures~\cite{angelini_criptopias_2023}. Autistic children use storytelling to engage in imaginative play~\cite{tartaro_storytelling_2006}. Individuals with visual impairments use storytelling to foster imagination, emotional growth, and language skills~\cite{chopra_storybox_2022}. Story consumers with blindness or low vision also use video stories to learn, entertain, and imagine~\cite{jiang_its_2024}.

\par
Recent HCI and accessibility research on storytelling has primarily focused on how technologies assist PwDs in everyday recording~\cite{hibbard_vlogging_2011, yoo_understanding_2021} and creative production~\cite{yoo_remembering_2024, chheda-kothary_engaging_2024, chopra_storybox_2022}. Generative AI introduces a new design material~\cite{yildirim_how_2022} for storytelling by automatically generating narratives and multimodal outputs, thereby reducing creative effort. A gap remains in understanding how GenAI can lower production barriers, scaffold narrative development, and enable self-expression for individuals who might otherwise struggle to share their stories. Through the lens of video-based disability stories, this research examines how PwDs unfamiliar with GenAI-based video creation engage with GenAI as an expressive tool, to inform the meaningful integration of GenAI into storytelling technologies for PwDs.

\subsection{Generative AI for People With Disabilities}
Generative AI (GenAI) refers to \textit{``artificial intelligence systems that can create new content, such as text, images, audio, or video, rather than just analyzing or acting on existing data''} \cite{gozalo-brizuela_chatgpt_2023}. Large language models (LLMs; e.g., ChatGPT, Gemini), image generators (e.g., DALL-E), and AI voiceover applications (e.g., ElevenLabs) have been rapidly adopted to support creative work~\cite{hu_designing_2025}. While GenAI offers potential benefits for PwDs, it also poses distinct risks.
\par
GenAI's content creation capabilities have supported the creative expression and productivity of PwDs, especially for individuals who lack media production skills. HCI research has shown that GenAI helps PwD creators engage in artistic practices by reducing media production barriers~\cite{bennett_painting_2024, saha_tutoria11y_2023} and enhancing independence~\cite{choi_exploring_2025, lee_altcanvas_2024}. GenAI may enable the creation of personalized and authentic content while enhancing productivity for personal expression~\cite{glazko_autoethnographic_2023, bennett_painting_2024, mcnally_disability_2024, choi_exploring_2025, lee_altcanvas_2024, s_guedes_artistic_2024}. Moreover, GenAI writing tools assist PwDs in navigating ableism, developing strategies to mitigate harm, and improving emotional well-being~\cite{mcnally_disability_2024}. While GenAI demonstrates promise in supporting self-expression, its role in facilitating the assembly of different story components has rarely been explored. Storytelling videos involve ideation, scripting, production, and post-production~\cite{choi_creator-friendly_2023, kim_unlocking_2024}, which may introduce new requirements for sustaining the benefits of GenAI in creative work.
\par
HCI research examining PwDs' use of GenAI has also focused on the challenges of GenAI in creative work. For example, AI-generated images may misrepresent PwDs through reductive, archetypal, or disrespectful portrayals~\cite{mack_they_2024, glazko_autoethnographic_2023} and often require verification of outputs~\cite{glazko_autoethnographic_2023, bennett_painting_2024, tang_everyday_2025}. GenAI outputs may misalign with creative intent and nuanced experiences~\cite{bennett_painting_2024, mcnally_disability_2024}, or introduce unwanted content~\cite{bennett_painting_2024, lee_altcanvas_2024}. While LLM outputs are not typically offensive or overtly toxic, PwDs have noted that GenAI-generated texts frequently perpetuate ableist stereotypes and biases~\cite{s_guedes_artistic_2024}, echoing tropes such as ``inspiration'' or ``disability porn''~\cite{gadiraju_i_2023, gogolushko_ai_2022, bennett_painting_2024}. Despite recognition of these challenges, further investigation is needed to understand how they affect PwDs' intentions to create story videos and how such challenges are manifested within the specific context of storytelling.
\par
While addressing GenAI representation issues is critical, GenAI-facilitated storytelling requires closer attention to how users articulate narratives and how unsuitable content may emerge. GenAI media lacking meaningful value is often perceived as lower quality~\cite{rae_effects_2024}, prompting social media platforms to restrict AI use~\cite{lloyd_personal_2019} and implement content labels~\cite{gamage_labeling_2025}. Understanding the motivations and needs of PwDs in GenAI-based video creation is essential to ensure such restrictions do not inadvertently hinder disability storytelling.

\subsection{Disability Storytelling through Video-Making}
In this research, we use video-making as a context to examine how PwDs instruct GenAI to create disability stories. Videos are a widely used medium for capturing and presenting such stories~\cite{rodger_journeycam_2019, jiang_its_2024, hibbard_vlogging_2011}, and they are commonly shared on social media to express identity and challenge disability stereotypes~\cite{duval_chasing_2021, cocq_self-representations_2020, struck-peregonczyk_changing_2023}. Storytelling videos educate non-disabled audiences by raising awareness and combating stigma~\cite{niu_please_2024, seo_understanding_2018, cocq_self-representations_2020}. 
\par
Our research is guided by Lambert's theory of \textit{digital storytelling}~\cite{lambert_digital_2013}, a seven-step framework for narrating personal stories. Accessibility research using this framework has demonstrated that video-based storytelling is a common and critical activity for PwDs, enabling them to engage in digital expression~\cite{saridaki_digital_2018}, preserve oral history~\cite{manning_original_2010}, and develop communication and leadership skills~\cite{sitter_building_2024}. Drawing on this framework, we examine three research questions concerning depicted moments and motivations, GenAI-supported story components, and intentions to share GenAI-generated videos.

\par
Creating stories requires PwDs to \textit{``own [their] insights''}~\cite{lambert_digital_2013}, a step in which storytellers decide not only what story to tell but also why and what deeper meanings they wish to convey. On platforms such as TikTok, PwD creators share their insights through playful skits, gamified therapy sessions, and stereotype-debunking videos that challenge misconceptions~\cite{duval_chasing_2021}. YouTube storytelling is often motivated by advocacy and creative expression~\cite{bromley_broadcasting_2008, niu_please_2024}. While GenAI can generate plausible media, it remains unclear what types of stories PwDs would be motivated to create with it.
\par
Digital storytelling also requires \textit{``owning your emotions''} and \textit{``finding the moment''}~\cite{lambert_digital_2013}, where storytellers identify the emotions present in their stories and select key moments that best convey their insights. However, the effectiveness of GenAI in capturing emotions and illustrating such moments in disability stories remains unknown. Creating video stories further involves \textit{``seeing''} and \textit{``hearing [the] story''}\cite{lambert_digital_2013}, which requires generating visuals and audio to accompany the narration of the moments. PwDs often face accessibility and skill-related barriers in this process~\cite{choi_its_2022, mcclimens_presentation_2008, sannon_disability_2023}. While GenAI automates media creation, it remains unclear what types of visual and audio materials PwDs would produce in storytelling. Moreover, GenAI content may portray PwDs in discriminatory or stereotypical ways~\cite{mack_they_2024, glazko_autoethnographic_2023}, making it essential to examine how PwDs navigate these challenges during story creation.
\par
Finally, storytelling requires \textit{``assembling''} and \textit{``sharing''}~\cite{lambert_digital_2013}, which involve compiling story elements into a complete video and deciding on the audience and mode of distribution. While GenAI-generated scripts and images allow PwDs to share stories without appearing on camera, their impact on sharing intentions remains unknown.

\section{Method}

\subsection{Study Context}


Our participants were nine PwD members from [ORGANIZATION NAME]\footnote{\label{note1}Hidden for anonymity} (see \autoref{tab:participants}), a non-profit organization that provides services to children and adults with disabilities. They are members of the [CAMPAIGN NAME]\footref{note1} campaign, which promotes disability history education through social media videos and public seminars. Participants had low experience creating short-form disability advocacy videos and delivering public speeches about disability history, but none had prior experience using GenAI for video production.

\par
\begin{figure}[!h]
    \centering
    \includegraphics[width=\linewidth]{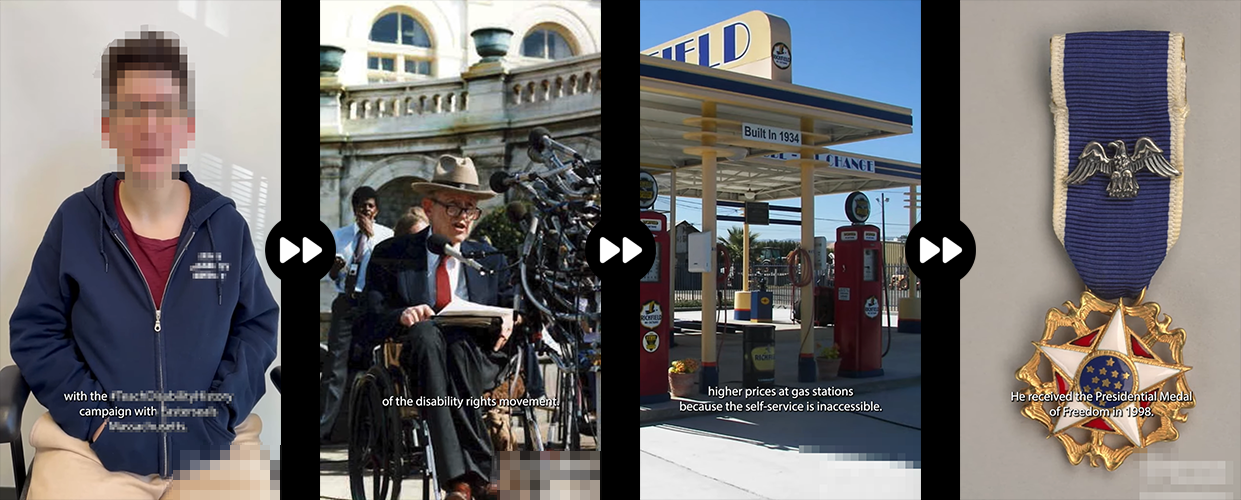}
    \caption{An example YouTube Short video produced by [CAMPAIGN NAME] in 2024. The video features a campaign member narrating a historical story about how Justin Dart Jr. advocated for PwDs to avoid paying higher prices at gas stations, accompanied by five static historical images illustrating the narration.}
    \Description{Four images side by side with arrows pointing from left to right between them. On the far left: a light-skinned person with their face blurred wearing a blue hoodie and tan pants sits in a chair facing the camera. There is text on the bottom of the image with some words blurred and unreadable. The visible portion says: “with the…campaign with…”. On the middle left: a light-skinned adult man wearing a tan hat, black glasses, a black suit, a red tie, and brown shoes faces towards the camera. He uses a wheelchair and is holding papers while speaking into several microphones. There is text on the bottom of the image that says “of the disability rights movement”. On the middle right: a gas station with a sign that says “built in 1934”. There is text on the bottom of the image that says “higher prices at gas stations because the self-service is inaccessible". On the far right: a medal with a red, white, and blue star surrounded by golden eagles on a blue and white ribbon with a silver eagle clasp. There is text on the bottom of the image that says “He received the Presidential Medal of Freedom in 1998”.}
    \label{fig:eastseals}
\end{figure}

In 2024, the [CAMPAIGN NAME] campaign produced 10 short-form videos (\autoref{fig:eastseals}) highlighting key events in disability history as part of its efforts to raise public awareness. Each video features a campaign member narrating a significant historical event related to disability, accompanied by 4–5 static historical images inserted throughout the narration. The videos last between 60 and 90 seconds. Each video was created by one PwD member who prepared the materials and narrated the script, while a technical staff member handled the filming and video editing. This study was initiated to explore whether GenAI can serve as an alternative pathway to enhance [CAMPAIGN NAME]'s social media presence. Following the [CAMPAIGN NAME] video style, we designed the GenAI video creation task (see \ref{subsubsec:creation_task}).
\par
Although the participant group was small and relatively homogeneous in disability representation, the [CAMPAIGN NAME] members brought valuable experience in public discourse on disability awareness through storytelling -- central to our research aims. Aligning with prior HCI and ASSETS studies~\cite{choi_its_2022, bennett_painting_2024, chheda-kothary_engaging_2024}, studying a dedicated PwD group enabled us to obtain reflective narrations that reveal provisional opportunities and challenges of adopting GenAI for disability storytelling, rather than general usage patterns across disability groups. While participants had limited prior experience with video production and GenAI media generation tools, engaging them offered a unique lens to examine how GenAI can serve as an accessible entry point for individuals who wish to share disability stories but face creative barriers~\cite{choi_its_2022, xiao_understanding_2025}.

\begin{table*}[!h]
    \centering
    \begin{tabular}{|p{0.05\textwidth}|p{0.05\textwidth}|p{0.08\textwidth}|p{0.3\textwidth}|p{0.16\textwidth}|p{0.32\textwidth}|}
    \hline
        ID & Age & Gender & Disability & GenAI Experience & Social Media Experience \\
    \hline
        P1 & 25-34 & Male & Genetic condition with symptoms similar to cerebral palsy & ChatGPT, CoPilot & FaceBook and Instagram \\
    \hline
        P2 & 25-34 & Female & Intellectual & Zona & Facebook, Instagram\\
    \hline
        P3 & 25-34 & Male & Cerebral Palsy & ChatGPT & Facebook and Instagram\\
    \hline
        P4 & 18-24 & Male & Autism Spectrum Disorder & ChatGPT & Facebook, YouTube, TikTok\\
    \hline
        P5 & 25-34 & Female & Globally delayed in all areas, low muscle tone, AP scoliosis & None & Facebook, Instagram\\
    \hline
        P6 & 18-24 & Male & A physical disability and an intellectual disability & Text to Speech & Instagram, TikTok\\
    \hline
        P7 & 25-34 & Female & ADD, Autism, and Epilepsy & None & Facebook, Instagram, and TikTok \\
    \hline
        P8 & 18-24 & Male & High-functioning autism and an intellectual disability & ChatGPT & Facebook \\
    \hline
        P9 & 25-34 & Male & Visually Impaired & Gemini, SeeingAI & None \\
    \hline
    \end{tabular}
    \caption{Participant Demographics}
    \label{tab:participants}
    \Description{The table summarizes demographics of nine participants aged 18-34. Three are female, six are male. There is a mix of intellectual and physical disabilities as well as varying levels of Generative AI and Social Media experience.}
\end{table*}

\subsection{Study Procedure}
The study was conducted over Zoom, where the researcher facilitated video creation with each participant by screen sharing GenAI tools, including ChatGPT (GPT-4o) for script generation, DALL-E 2 for image creation, ElevenLabs for AI voiceover, and CapCut for video editing. To qualify, participants confirmed they were at least 18 and had at least one disability. Each 1.5-hour session consisted of a 30-minute pre-study, $\sim$40 minutes of video creation, and a $\sim$20-minute post-study survey. Participants received a \$50 gift card as compensation. The study data used for qualitative analysis consisted of anonymized video recording transcripts and the prompts used during the study. The pre-study survey included a task comprehension check and the consent form. The study was approved by the IRB at the authors' institution.


\begin{figure}[!h]
    \centering
    \includegraphics[width=\linewidth]{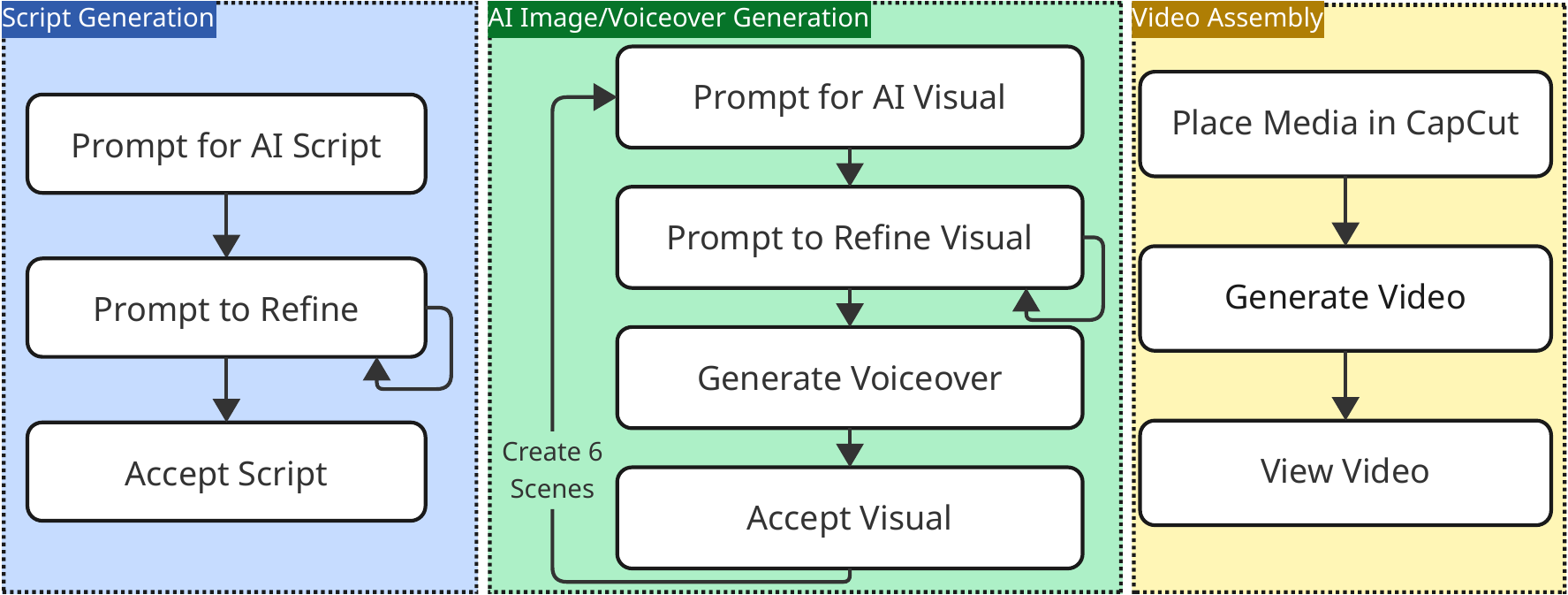}
    \caption{Creation Task Flow}
    \label{fig:FlowChart}
    \Description[A flow chart depicting three creation tasks: Script Generation, AI Image/Voiceover Generation, and Video Assembly]{Script Generation includes Prompt for AI Script, then Prompt to Refine, then either Prompt to Refine again or Accept Script. AI Image/Voiceover Generation includes Prompt for AI Visual, then Prompt to Refine Visual, then either Prompt to Refine Visual again or Generate Voiceover, after Generate Voiceover is Accept Visual, this repeats six times to make six scenes. Video Assembly includes Place Media in CapCut, Generate Video, then View Video.}
\end{figure}

\subsubsection{Pre-study Session}
The pre-study session introduced the study procedure and demonstrated how to use GenAI tools to create a script, generate images and voiceovers, and produce a final demo video. The demonstration was an example video about \textit{``A wheelchair user is frustrated when an Uber driver cannot give them a ride.''} The researcher showed how to generate and edit a script with ChatGPT, create images for the first two scenes, produce voiceovers, and combine them into an output video. During early contact, the campaign organizer informed us that all members ``can use computers and no disability conditions would prevent them from reading or speaking.'' During this session, the researcher asked the participant to summarize the purpose and steps of the study and to indicate whether they felt confident providing GenAI instructions, in order to validate their eligibility. We confirmed participants' understanding when they explicitly stated that they would ``make a video'' and ``use AI'' and that they were confident in providing GenAI instructions to the researcher. Once participants demonstrated they could participate, the researcher obtained their signed consent form. Finally, demographic questions were asked, including age range, gender, disability, experience with generative AI, and experience posting on social media.

\subsubsection{Creation Tasks}
\label{subsubsec:creation_task}
We adapted the [CAMPAIGN NAME] video format (\autoref{fig:eastseals}) to design the task. Each GenAI video was $\sim$1- minute long, with an AI-generated voiceover and six DALL-E images, each paired with $\sim$10-second of ChatGPT-generated narration. The GenAI videos mirrored the pacing and structure of the [CAMPAIGN NAME] videos, in which static images are used to illustrate narrated content. Participants were instructed: \textit{``Use the tool to make a video. Come up with a story you want to tell. The main character does not have to be you; it can be anyone. The story must be disability-related.''} 
\par
This task design allowed us to observe the stories they chose to create with GenAI and assess the topic, emotions, context, and quality of the GenAI content. The six-scene format was determined through pilot tests with non-disabled users, which showed that creating scripts and generating images took about 25 minutes. Standardizing the number of scenes ensured consistent session lengths and reduced fatigue from generation delays. We did not ask participants to create videos about disability history due to inaccuracies in GenAI-produced historical imagery~\cite{kaur_is_2025}. We also avoided AI talking heads or GenAI-produced animations because longer rendering times (e.g., 2–3 minutes for an 8-second video clip) would have limited PwDs' ability to iterate GenAI outputs during the study.
\par
The video production involved three stages: script generation, AI image/voiceover generation, and video assembly (\autoref{fig:FlowChart}). These steps were derived from the common video production pipeline~\cite{kim_unlocking_2024,anderson_making_2025}. Despite the limited variety of GenAI tools and video formats used, our procedure reflected essential steps of GenAI-enhanced YouTube video creation, including planning scripts with an LLM, preparing visual materials, and iteratively refining prompts to improve content quality~\cite{anderson_making_2025, kim_unlocking_2024}. Because the study focused on the outcomes of GenAI content rather than the usability of GenAI software or video editing tools, the researcher provided procedural guidance and operated the GenAI tools. Participants read aloud or typed out the desired prompts, and the researcher entered them into ChatGPT. This strategy reduced the barrier of typing long prompts, enabled participants to concentrate on crafting the story and evaluating the GenAI output, and allowed the task to be completed within a reasonable time frame. One participant (P9) had a visual impairment. The researcher read aloud each script version. Image descriptions were generated by ChatGPT with the prompt: \textit{``make a description of the image''} and were read aloud by the researcher. After the video was completed, the researcher played the video with voice over.
\par
For \textbf{script generation}, each script included six scenes, with each scene containing a story narration. The researcher asked participants to complete a prompt template: \textit{``I want to create a 1-minute narrated story script about} \texttt{[PARTICIPANT INPUT]} \textit{. I plan to show six different scenes in the video, with the narration of each scene taking 10 seconds to read. Generate timestamps for each scene along with the script.''} Participants described the disability story they wished to create and refined the prompt until they were satisfied with the video story script. An example output from ChatGPT is:

\begin{center}
    \label{box:example_output}
    \shadowbox{
        \begin{minipage}{\dimexpr1\linewidth-2\fboxsep-2\fboxrule}
Title: "[NAME]'s Lunch Outing – Denied by Stairs"\\
Scene 1 — [0:00–0:10]\\
``[NAME] walks slowly down the sidewalk using his walker. He stops and looks up at a restaurant he's been wanting to try for weeks.\\
Narration:\\
``This is [NAME]. It's lunchtime, and he's been craving this spot for days. He finally made the trip, walker and all.''\\
...\\
Scene 6 — [0:50–1:00]\\
Final shot shows the stairs again, then fades to black with text: ``Accessibility is not optional.''\\
Narration:\\
``These stairs are more than steps — they're walls. Until we break them down, people like [NAME] will keep being left out.''
        \end{minipage}
    }
\end{center}

To prompt participants in developing their stories, the researcher offered guiding questions such as: ``What disability would you like to highlight?'' ``What challenge or situation would you like to focus on?'' ``Where does it take place?'' ``Would you like the story to be about you, or someone else?'' and ``Are there any specific details you want to include?'' After creating the initial script, participants had the opportunity to request revisions from ChatGPT. Once the script was accepted, the next step was image generation.
\par

For \textbf{AI image/voiceover generation}, participants created six images, one for each scene generated by ChatGPT. The following prompt template was used: \textit{``Create an image of} \texttt{[PARTICIPANT INPUT]}\textit{.''} If participants were unsatisfied with an image, they could ask ChatGPT to revise or regenerate it. The script for each corresponding narration was then converted into a voiceover using ElevenLabs.
\par

For \textbf{video assembly}, once a participant accepted the six images and voiceovers, the researcher used CapCut to assemble the video. Each image was matched to its corresponding voiceover. Participants then watched their final video. 
\par

\subsubsection{Post-study Interview}
During the post-study interview, participants responded to a series of semi-structured questions derived from Lambert's digital storytelling theory~\cite{lambert_digital_2013} to address the three research questions. RQ1 examines how participants used GenAI to take ownership of story insights, focusing on their motivations for creating disability stories. RQ2 investigates how GenAI media supported participants in creating the storyline and expressing emotions. RQ3 explores participants' perceptions of sharing GenAI-created disability storytelling videos.

\subsection{Data Analysis}
Our examination of GenAI's roles draws on two data sources: participants' prompts to ChatGPT and their semi-structured interview responses. The interview questions were asked in an objective manner (e.g., for emotion, \textit{``How do you feel about GenAI's ability to accurately understand and convey the emotional tone you intend to express?''}). During the task, ChatGPT conversations were saved, including all inputs and the corresponding text and image outputs, and these logs were used to address RQ2. 
\par
The researchers conducted a thematic analysis~\cite{braun_thematic_2019} of the interview data and GenAI prompts. In the open coding stage, interview responses and ChatGPT prompts were converted into digital cards, and three researchers reflected together on how they related to story components, GenAI activities, and sharing motivations. During axial coding, we used affinity diagramming to iteratively group quotes for RQ1 and RQ3 around emerging themes. For RQ2, we first grouped the prompts into themes and then added relevant interview quotes. In the selective coding phase, all quotes and prompts were split into semantic units (one or a few individually interpretable sentences). Two researchers independently annotated all units using the identified themes, achieving substantial agreement (Krippendorff's $\alpha$ with Jaccard metric: 0.83 for RQ1; 0.92 for interview feedback and 0.78 for ChatGPT prompts in RQ2; and 0.99 for RQ3). A third researcher reviewed and resolved remaining discrepancies.

\section{Results}
\subsection{RQ1: Story Moments and Motivation}
RQ1 examines the potential motivations of PwDs for using GenAI to create disability stories and the types of stories they aim to produce with ChatGPT. During the interview, we asked whether participants had any preference for the type of story they wanted to create with ChatGPT and whether AI tools helped them depict the story.
\begin{figure*}[!h]
\centering
    \begin{tabular}{lll}
        \begin{subfigure}[t]{.32\textwidth}
        \includegraphics[width=\textwidth]{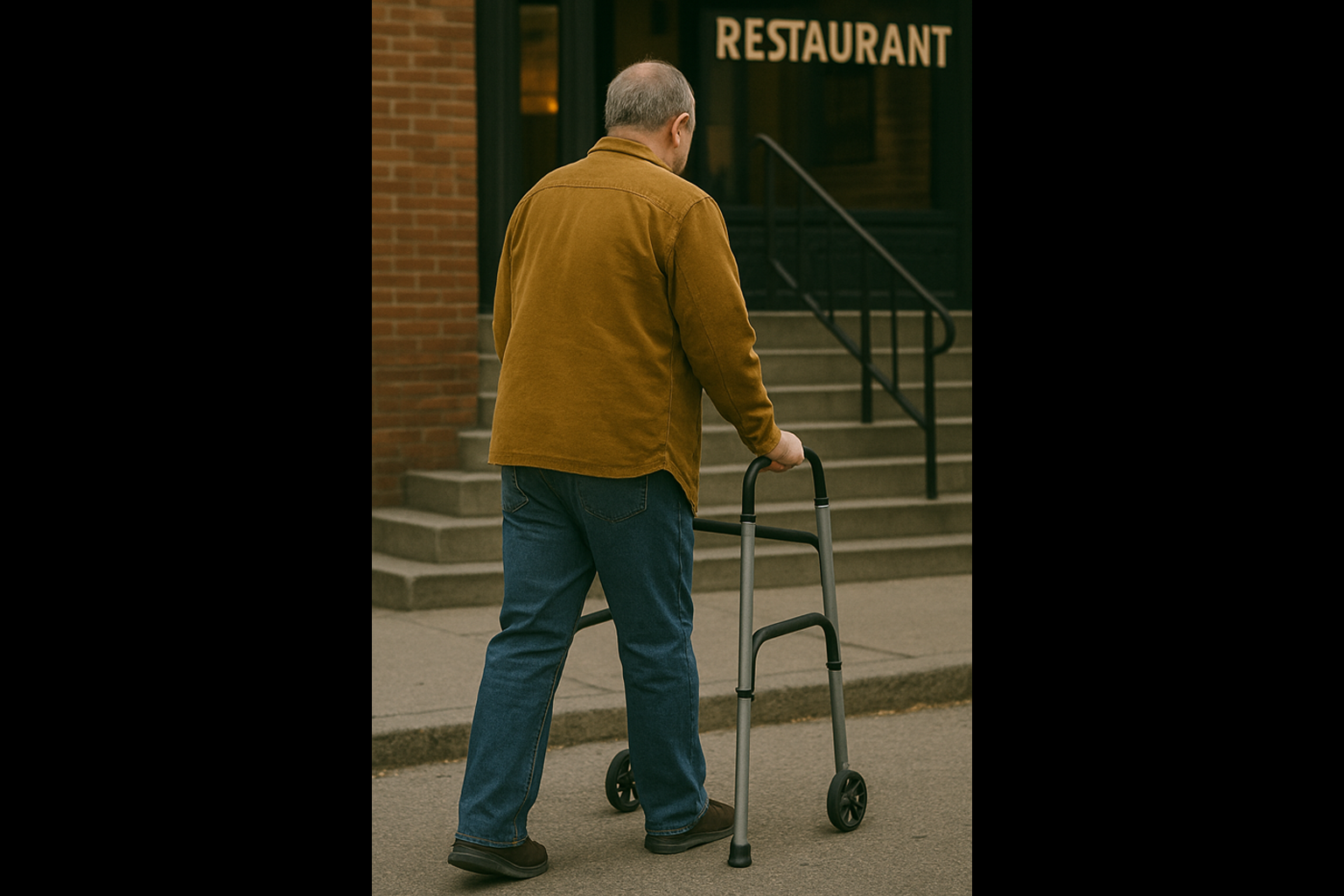}
        \caption{(P1) A man who uses a walker is unable to go to a restaurant as the stairs are not accessible.}
        \label{fig:P1}
        \Description{A light-skinned middle aged man with gray hair wearing blue jeans and a brown shirt uses an anterior walker as he walks away from the camera towards a Restaurant with stairs leading up to the door.}
        \end{subfigure}
        &
        \begin{subfigure}[t]{.32\textwidth}
        \includegraphics[width=\textwidth]{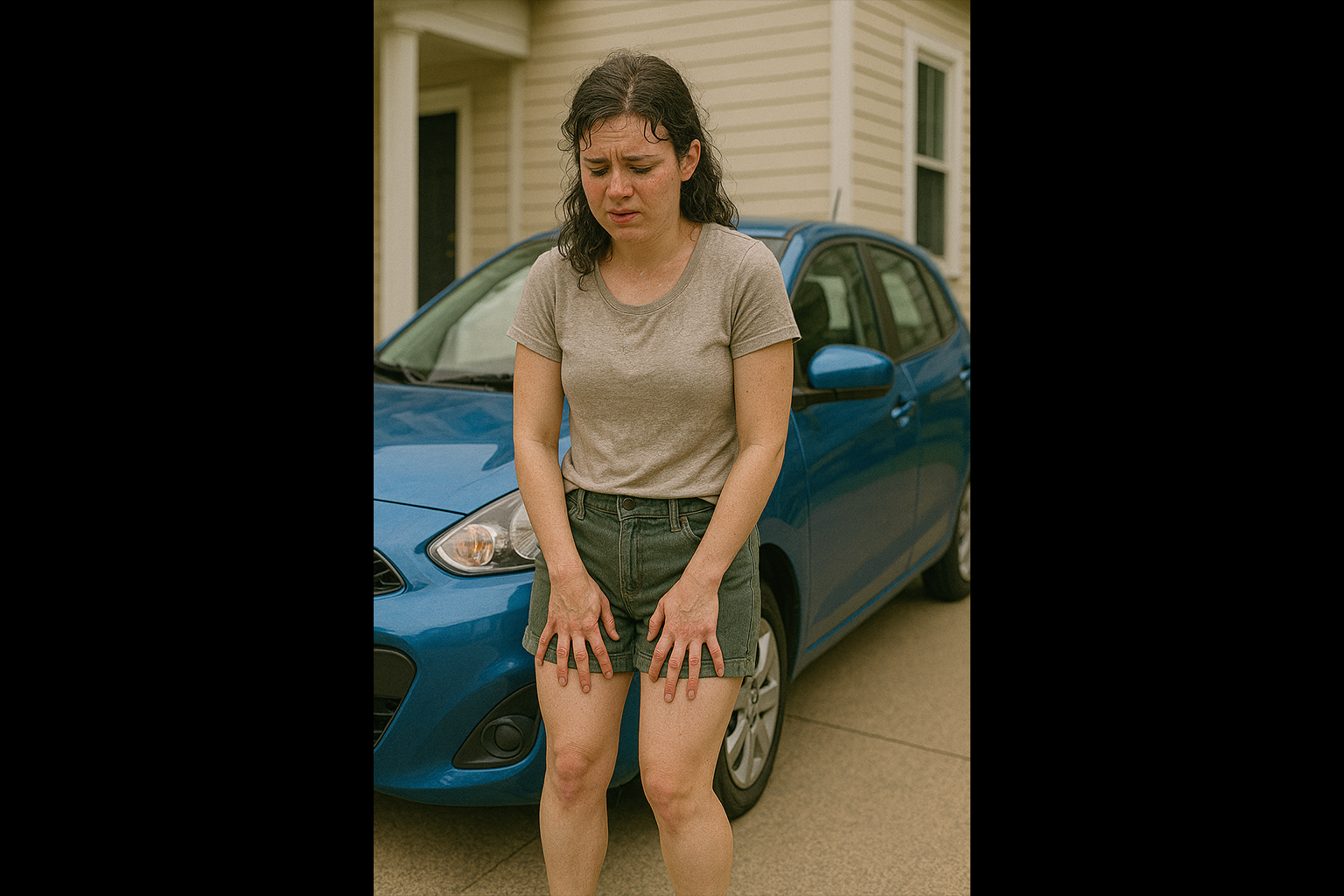}
        \caption{(P2) A woman who has Autism drives for the first time, overcoming her fear.}
        \label{fig:P2}
        \Description{A light-skinned adult woman with long brown hair wearing jean shorts and a grey short sleeved shirt faces the camera, looking down as she wipes her hands on her shorts. A blue car is parked behind her in front of a house.}
        \end{subfigure}
        &
        \begin{subfigure}[t]{.32\textwidth}
        \includegraphics[width=\textwidth]{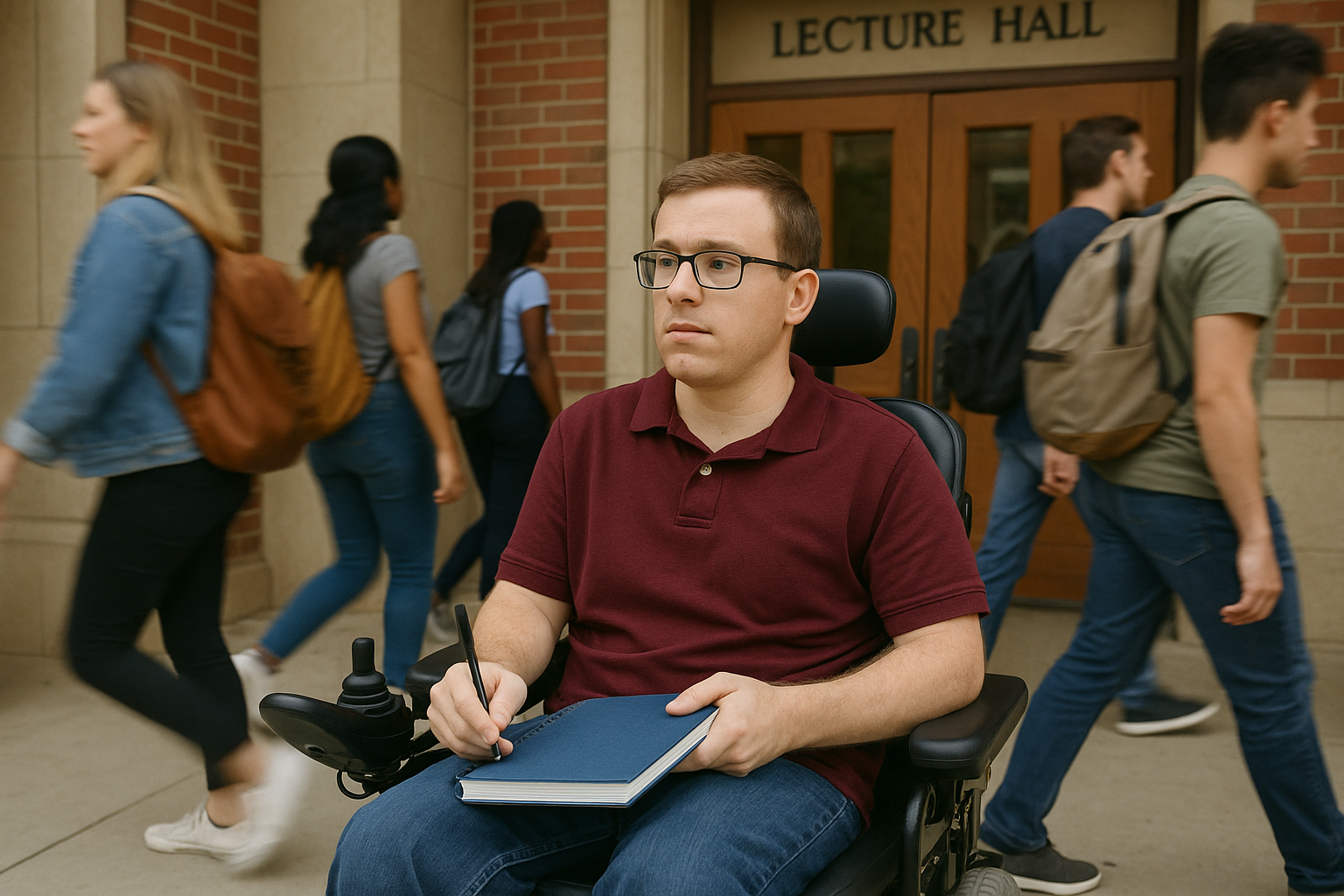}
        \caption{(P3) A man with Cerebral Palsy struggles to get help writing notes in class, when someone stops and helps.}
        \label{fig:P3}
        \Description{A light-skinned adult man with short brown hair wearing blue jeans and a red collared shirt using an electric wheelchair faces the camera outside a lecture hall. He is holding a black pen and a blue notebook as various students walk behind him in the background.}
        \end{subfigure}
        \\
        \begin{subfigure}[t]{.32\textwidth}
        \includegraphics[width=\textwidth]{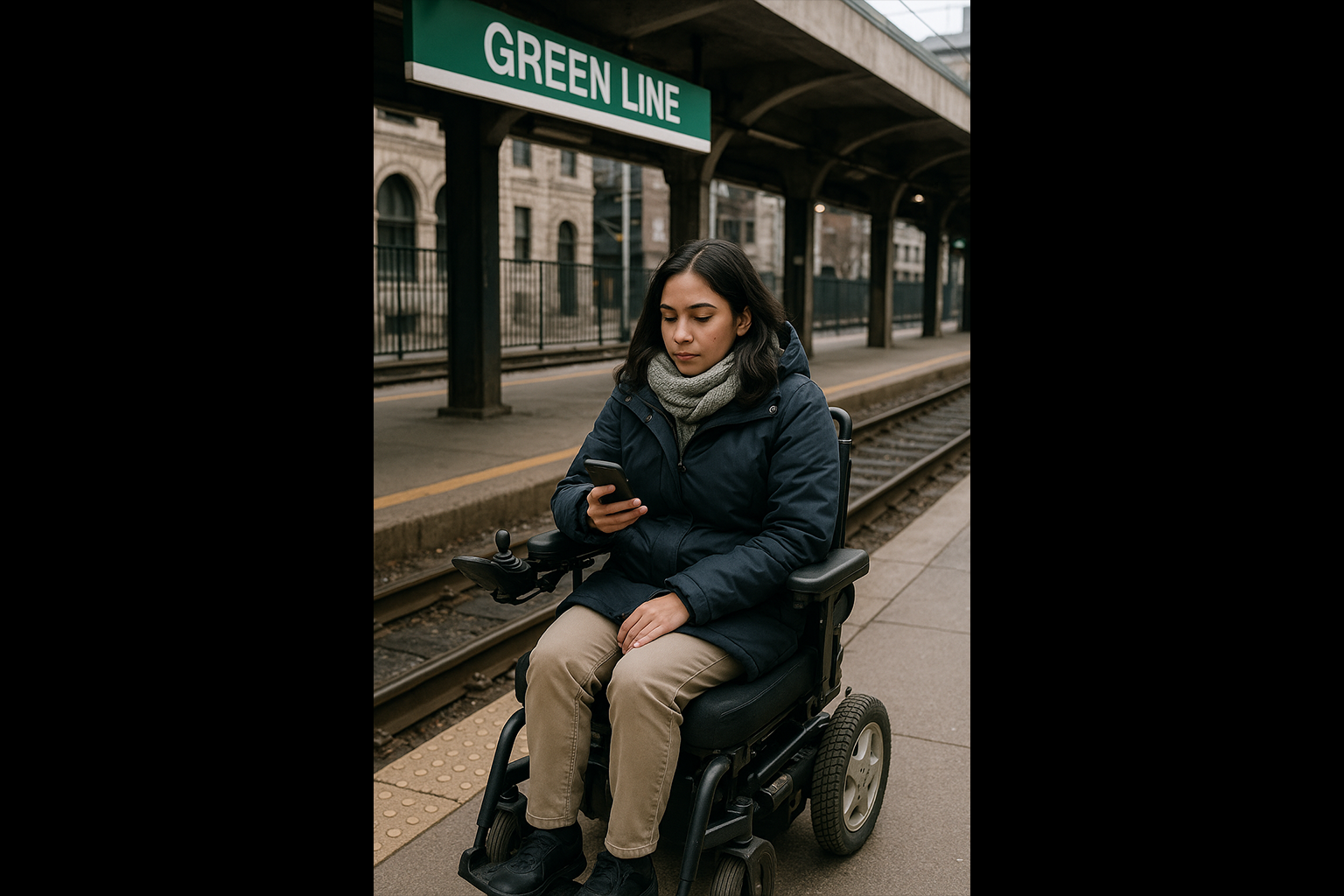}
        \caption{(P4) A wheelchair user is denied access to a green line train in Boston as the ramps are not accessible.}
        \label{fig:P4}
        \Description{A medium-skinned adult woman with shoulder-length dark hair wearing tan pants, a blue jacket, and a grey scarf using an electric wheelchair waits at a Green Line train station. She is facing the camera, looking down at her phone in her hand.}
        \end{subfigure}
        &
        \begin{subfigure}[t]{.32\textwidth}
        \includegraphics[width=\textwidth]{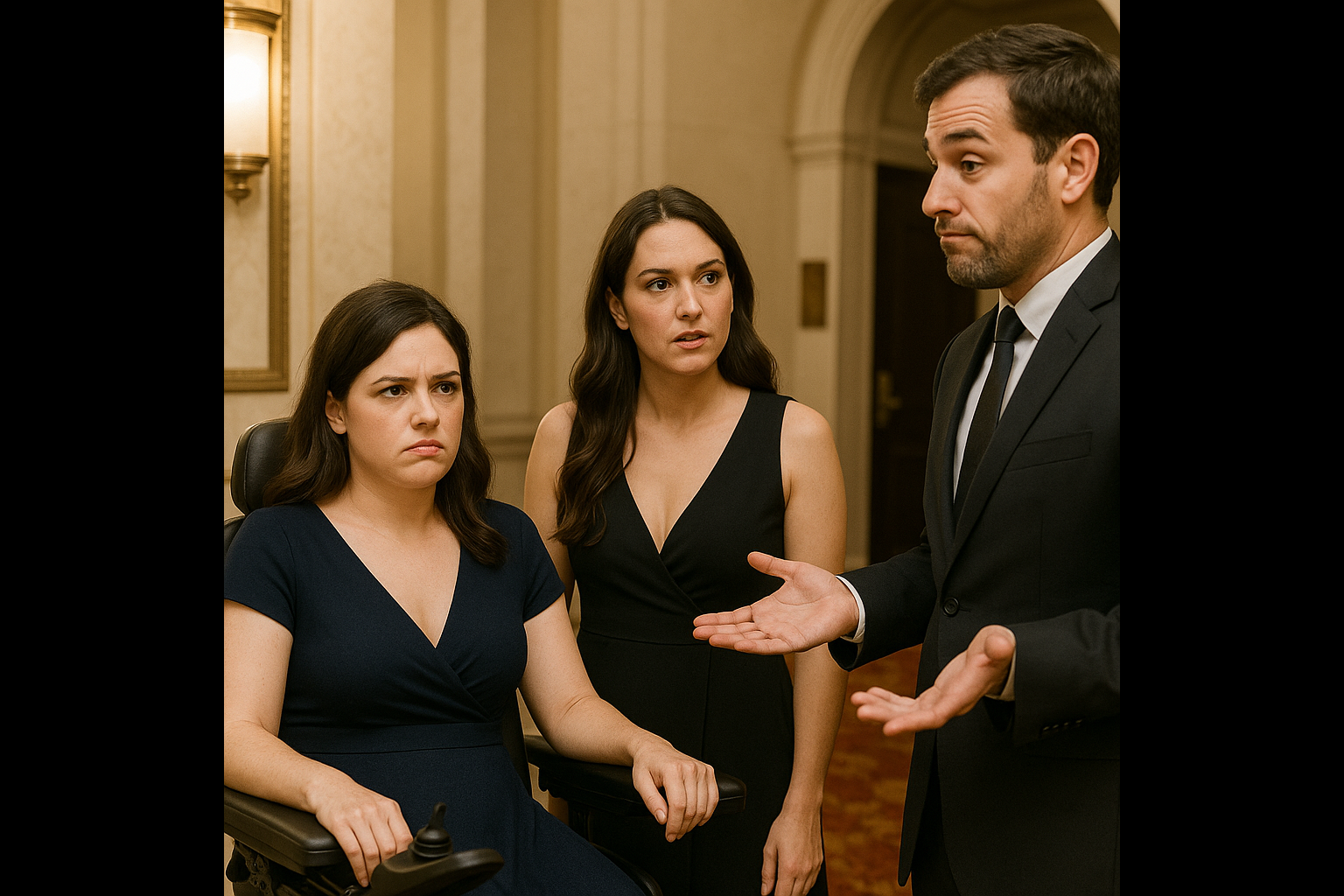}
        \caption{(P5) A wheelchair user is denied access to a show in a theater as there is no elevator to the balcony.}
        \label{fig:P5}
        \Description{Two adult women are side by side facing the camera. They are both smiling. On the left: a light-skinned woman with shoulder-length ginger hair. She wears a sparkly dark blue dress and uses an electric wheelchair. She is holding two tickets. On the right: a light-skinned woman with long dark hair hands a ticket to the woman on the left.}
        \end{subfigure}
        &
        \begin{subfigure}[t]{.32\textwidth}
        \includegraphics[width=\textwidth]{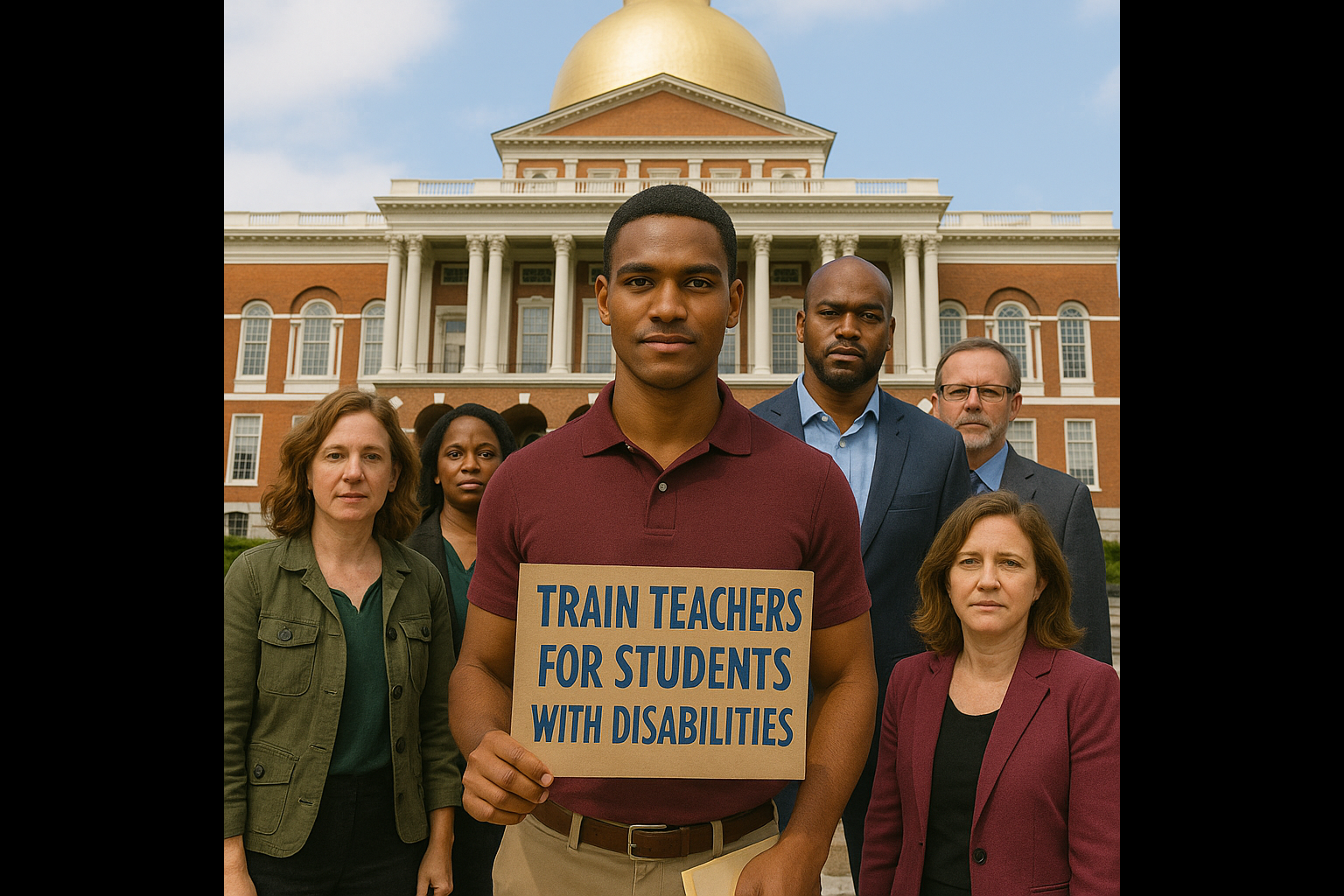}
        \caption{(P6) A disability advocate is denied the opportunity to speak at the Massachusetts State House.}
        \label{fig:P6}
        \Description{A dark-skinned adult man with short black hair faces the camera, holding a brown sign with blue text that reads "Train Teachers for Students with Disabilities". He is wearing a red collared-shirt and tan pants. Behind him is the Massachusetts State House and a crowd of five people standing with him.}
        \end{subfigure}
        \\
        \begin{subfigure}[t]{.32\textwidth}
        \includegraphics[width=\textwidth]{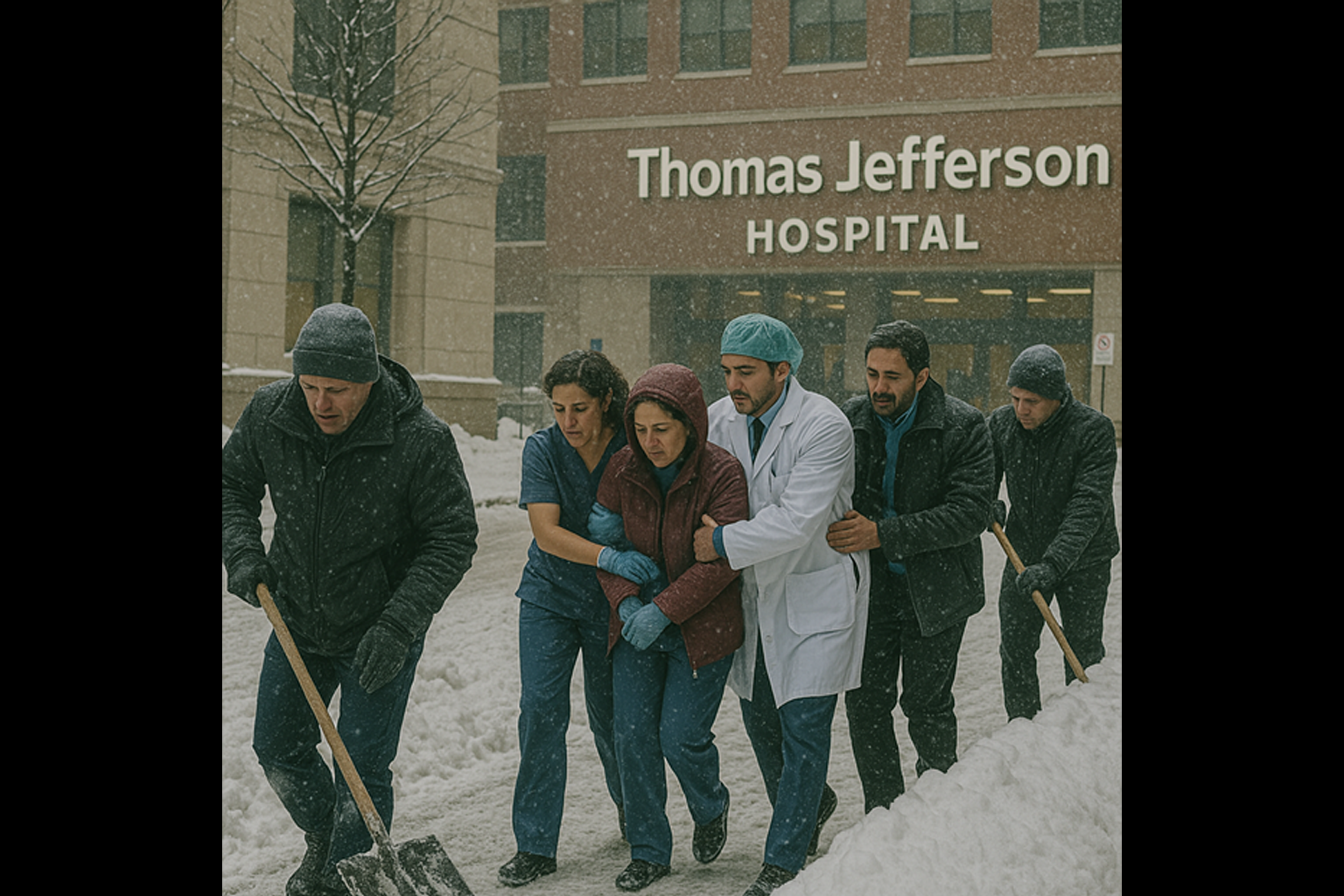}
        \caption{(P7) A group of people prone to seizures help people with mobility disabilities get inside a hospital during a winter storm.}
        \label{fig:P7}
        \Description{Six people face towards the camera, walking on a snow-covered road with Thomas Jefferson Hospital behind them. They all look down, with serious expressions on their faces as snow falls. From left to right: A light-skinned adult man wearing blue pants, a gray jacket, a winter hat, and gloves shovels a path. Next to him is a medium-light-skinned adult woman with long dark hair wearing blue nurse attire and medical gloves helps the person next to her walk through the snow. The person she is helping is a medium-light-skinned adult woman wearing a red coat with the hood up, blue pants, and medical gloves. Next to her is a medium-light-skinned adult man also helping her walk. He is wearing a white medical coat over a blue collared-shirt, a medical cap, and blue pants. Next to him is a medium-light-skinned adult man wearing a blue shirt, a gray jacket, and gray pants holding onto the man to his left's coat. Next to him is a light-skinned adult man wearing gray pants, a gray jacket, a winter hat, and gloves shoveling snow off the road.}
        \end{subfigure}
        &
        \begin{subfigure}[t]{.32\textwidth}
        \includegraphics[width=\textwidth]{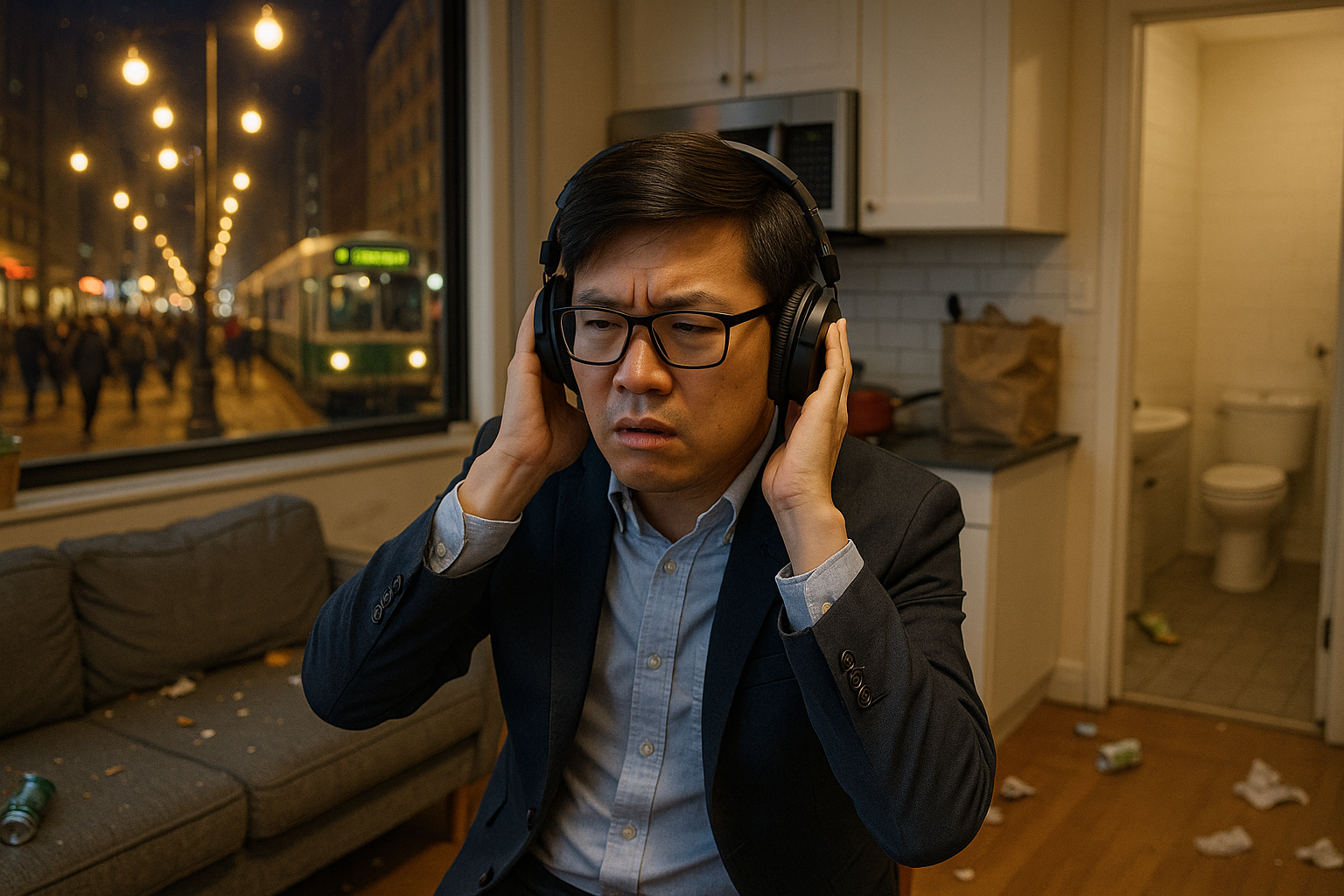}
        \caption{(P8) A man with Autism navigates sensory stresses, social struggles, and adult life with community, faith, and small routines. }
        \label{fig:P8}
        \Description{A medium-skinned adult man wearing a blue collared shirt, a dark blue blazer, and black over-ear headphones faces the camera. A messy apartment is behind him, with cans and food on the floor and couch. Street lights, a train, and crowds of people are outside a window to his left. He has a concerned look on his face and holds his hands over the headphones.}
        \end{subfigure}
        &
        \begin{subfigure}[t]{.32\textwidth}
        \includegraphics[width=\textwidth]{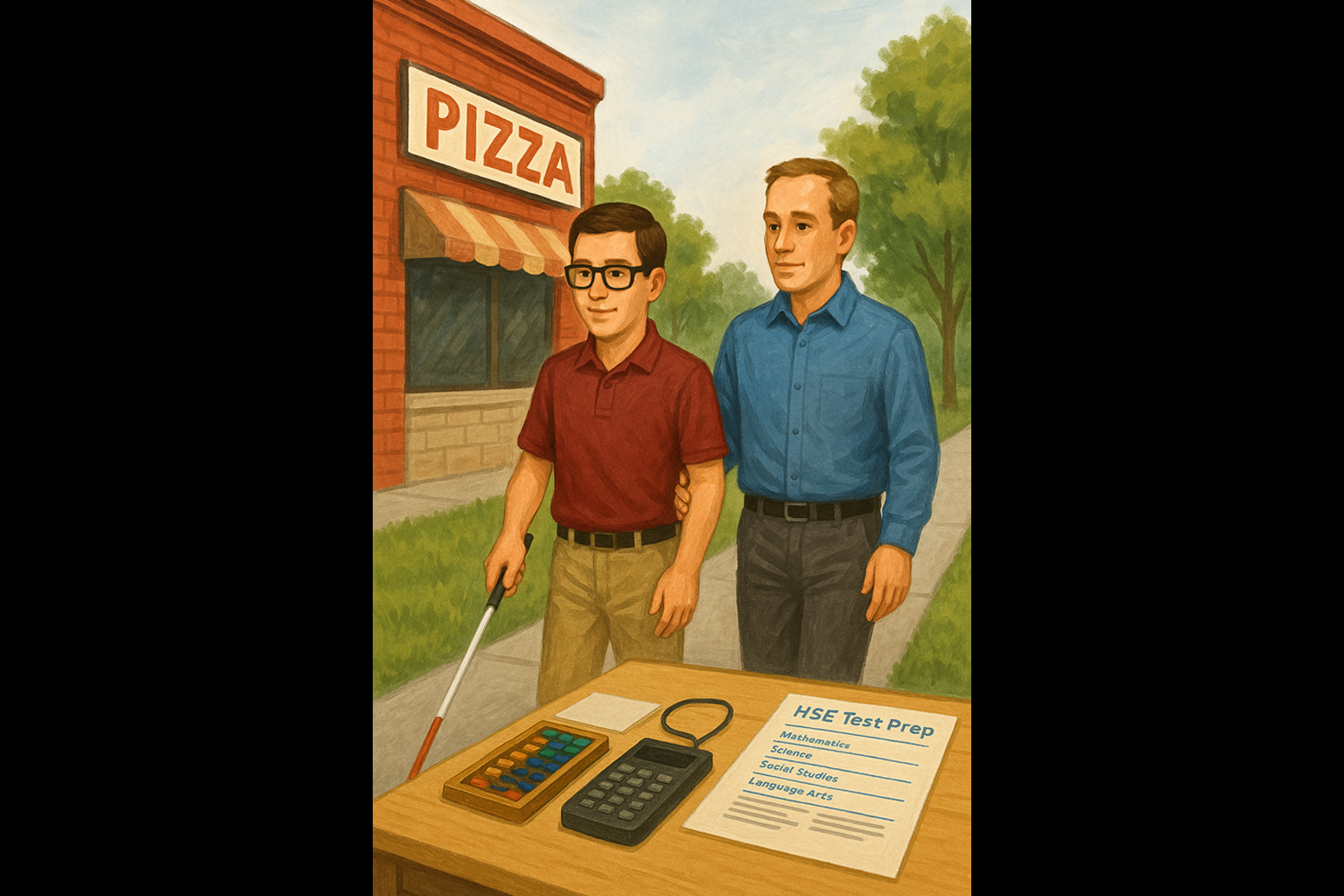}
        \caption{(P9) A man with visual impairment works towards his High School Equivalency Certification }
        \label{fig:P9}
        \Description{A drawing-style image of two light-skinned adult men walking side by side, a pizza shop behind them, and a table in front of them with an abacus, a calculator, and a paper on top of it. The paper has written on it: "HSE Test Prep" with sections titled "Mathematics", "Science", "Social Studies" and "Language Arts". From left to right: the man is wearing a red collared-shirt, tan pants, and black square glasses. He has short brown hair. He is using a white cane with a red bottom as he walks towards the camera. Next to him, the other man has short blonde hair, wearing a blue collared-shirt and gray pants as he holds the other man's arm.}
        \end{subfigure}
    \end{tabular}
\caption{Stories created by each participant and example scenes. The captions under each image summarize the stories created by the participant.}
\Description{A three by three grid of videos created by participants.}
\label{fig:all_story}
\end{figure*}

\subsubsection{Real-World Disability Challenges}
During the study, participants used GenAI to create videos illustrating real-world challenges related to inaccessible built environments (P1, P4, P5, P8), lack of social or societal support (P3, P6, P9), challenges caused by weather (P7), and mental health stress (P2 and P8). See \autoref{fig:all_story}. 
\par
In our interviews, six participants (P1, P2, P4, P5, P6, and P9) described how GenAI helped them illustrate real-world disability challenges. Participants emphasized that GenAI can transform everyday barriers into visual narratives. For example, P1 explained: \textit{``I feel like it's a common problem that people with disabilities go through and I think that could be very good if more people saw it.''} GenAI helped convey situations that the public often does not see. For example, P5 shared: \textit{``That way they can see a live action video... people don't usually see, people can have an idea what is going on in their community.''}

\subsubsection{Personal Disability Expression}
Participants used GenAI videos to express their opinions and feelings about disability (\autoref{fig:all_story}). In their videos, P1, P2, P3, P4, P5, P7, and P8 used GenAI to create stories based on lived experiences and to illustrate their frustration and feelings of helplessness. P6 and P9 expressed imaginative and wishful future activities they hoped to do. GenAI was also used by P3, P7, and P8 to express gratitude toward people who offered help. P2 and P8 used the videos to express their self-efficacy in overcoming mental health challenges.
\par
In the interview, all participants described how GenAI influenced the expression of disability stories. They felt that GenAI can support expression when they have difficulty articulating their thoughts. For example, P1 noted that \textit{``[GenAI] can give them sort of a voice. If [PwDs] have trouble, like speaking verbally.''} Participants also commented on how GenAI helps them reflect on disability stories. P8 mentioned: \textit{``ChatGPT helps narrate [the character's] life better and help me understand the struggle and where the downfall is.''} Participants appreciated the ability to use alternative identities in GenAI-generated stories to safely express personal experiences. Seven participants chose to use virtual characters rather than their real identities with ChatGPT, while two used their own names to construct narratives. P9 shared: \textit{``My name is not [NAME]. I am using my experiences and having the other characters go through what I go through''} and \textit{``[NAME] is the guy I want to be, telling disability advocacy stories.''}

\subsubsection{Disability Awareness}
In participants' videos, P6 and P9 called for government and major media outlets to raise political awareness and promote changes to laws or regulations to protect PwDs (\autoref{fig:all_story}). P1, P3, P4, P5, and P6 included call-to-action statements in their scripts (e.g., P1: \textit{``Accessibility is not optional''}; P4: \textit{``Accessibility is not a request, it is a right.''}) 
\par
During the interview, all participants described how GenAI videos can serve as a tool for raising disability awareness. Participants felt that GenAI helps make overlooked challenges more understandable to the public. For example, P1 mentioned that they hope the GenAI story can \textit{``give them a reason why businesses need to be made accessible.''} P4 wanted the MBTA in Boston to recognize that the ramp is too steep. P5 noted that people often assume wheelchair users already have access, but many bathrooms and doors are not accessible. Participants also emphasized that GenAI should be put to good use like empowering PwDs to advocate for and represent themselves. P6 noted, \textit{``I want to show that AI can still create good impact, as AI right now is a murky topic. But to show, hey, it can create good will.''}

\subsubsection{Disability Community Building}
In participants' videos, P3 and P7 used GenAI to illustrate scenes showing how others can help. P6's story depicted how the character sought to be a voice for PwDs in government. P8 highlighted that the church gave the character strength and hope. See \autoref{fig:all_story}.
\par
During the interviews, two participants, P2 and P7, specifically noted the motivation to use GenAI to support the exchange of disability experiences. For example, P7 stated, \textit{``I feel like that's good because [AI videos] help people with disabilities learn about other people with disabilities.''} Sharing stories through GenAI also helped P2 feel that such communication fosters a sense of solidarity, as they mentioned: \textit{``[Creating GenAI videos] helps people know that they are not alone. They have someone to connect with [similar] experiences.''}

\subsection{RQ2: Co-Creation of Disability Story Components with GenAI}
RQ2 examined participants' prompts and post-study feedback to understand how GenAI supports identity expression. We designed four semi-structured questions according to~\cite{lambert_digital_2013} regarding what participants liked or disliked about the GenAI stories, whether AI helped them gain new insights about disability, how GenAI materials represented disability identities, how they perceived the quality of the AI-generated video materials, and how GenAI conveyed emotional tones. We first categorized the prompts and then grouped interview responses into the identified GenAI creation themes. 

\subsubsection{Story Scripting and Ideation}
In our study, participants narrated the stories they wanted to tell, and ChatGPT produced the scripts. This theme includes the initial prompts used to generate the stories. For example, P1 created a script describing: \texttt{``Someone named [NAME] with a physical disability who uses a walker. They are trying to enter a restaurant and cannot because there are stairs and no elevator, there are lots of stairs and they would like to eat something but cannot because it is not ground level.''} Some participants asked ChatGPT to highlight disability identities. For example, P2 remade the script and specified that the character should \texttt{``get out of car, she should say ``look mom I was able to do this.''''} P6 informed ChatGPT to update the script so that the character \texttt{``came for his job ``Massachusetts Advocates for Children''''}.

\par
In the interview, participants suggested that GenAI improved spelling and wording when PwDs were not confident in their writing skills. For example, P2 noted that \textit{``[ChatGPT] helps make the story spelled right and everything.''} P4 similarly said, \textit{``I need to improve my writing. And so maybe AI helps you with this.''} Participants also described how GenAI facilitated story completion and enrichment. P2 stated: \textit{``I like that it is able to help us with a story. If we are able to describe what we want, it helps us get started.''} P4 further noted that GenAI helps them figure out what to write instead of stressing out and crumpling up drafts. 
\par
However, although identifying insights is a critical step in story scripting~\cite{lambert_digital_2013}, participants stressed that this should not be taken over by AI and that PwDs should retain autonomy in choosing their story ideas. P1, P7, P8, and P9 explicitly noted that they did not use GenAI to gain insights or identify the story moment. P3 further stated: \textit{``Unless it is the person creating the video, [GenAI] should not take the person's voice.''}

\subsubsection{Character Profiling and Picturing}
During the task, participants created and revised scenes to depict a character and define the story's setting. They actively described the character's actions, appearance, assistive technologies, emotions, and contextual environment.
\par
\begin{figure}[!h]
    \centering
    \includegraphics[width=1\linewidth]{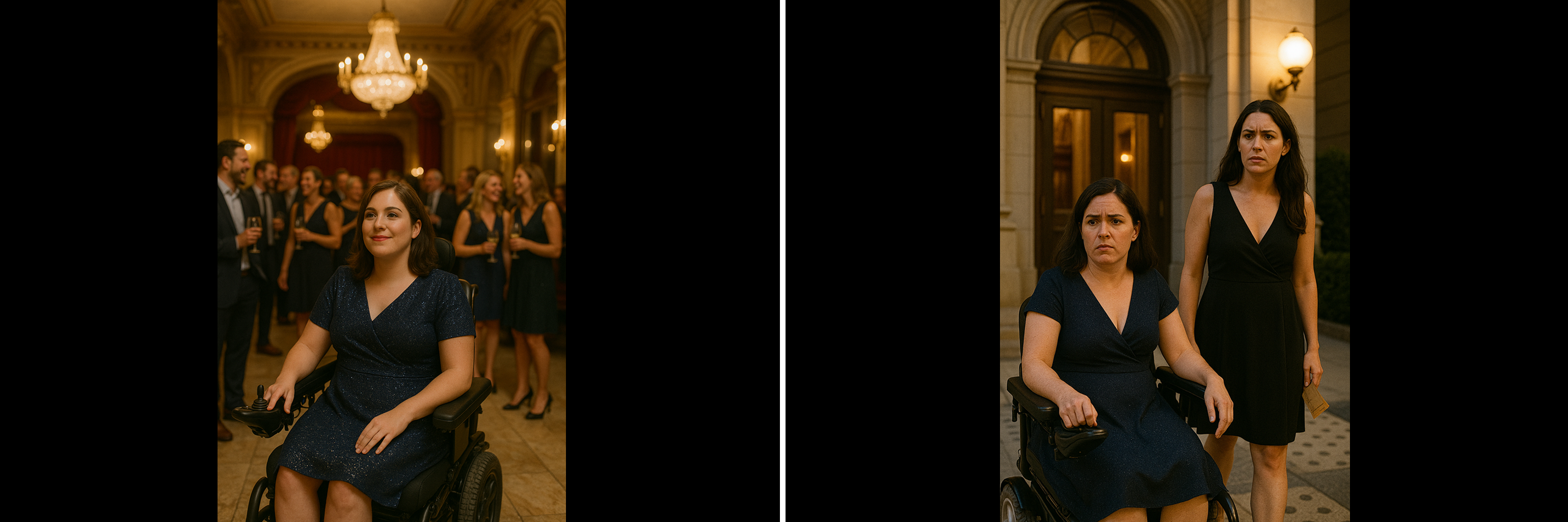}
    \caption{Left: P5 requested ChatGPT to depict people laughing and drinking in the theater. Right: P5 requested ChatGPT to illustrate that the PwD had to leave the theater.}
    \label{fig:P5_1}
    \Description{Two images side by side. On the left: a light-skinned woman with shoulder-length brown hair. She wears a sparkly dark blue dress and uses an electric wheelchair. She is facing the camera and smiling, inside of a theater with a crowd talking and drinking behind her. On the right: two adult women are side by side facing the camera. They are outside of the theater and both are frowning. On the left: a light-skinned woman with shoulder-length dark hair. She wears a sparkly dark blue dress and uses an electric wheelchair. On the right: a light-skinned woman with long dark hair holding a ticket.}
\end{figure}

\textbf{Action Presentation.} In their prompts, all participants described key actions for ChatGPT to build upon, and they used GenAI to visualize the feelings of characters. For example, in depicting a character's action, P2 prompted a script revision about a person with autism overcoming the fear of driving: \texttt{``In scene 2 as [NAME] clicks the seatbelt she should take a deep breath.''} To illustrate how a wheelchair user wanted to join others enjoying themselves in a theater but was ultimately denied because there was no elevator, P5 asked ChatGPT to remake several scenes with prompts such as \texttt{``the same but have people laughing and drinking''} and \texttt{``[the staff] has the [characters] leave the theatre''} (\autoref{fig:P5_1}). 
\par

In the interview, participants noted a benefit of GenAI in capturing emotional or physical actions without requiring risky real-life recording. P3 commented that \textit{``it can generate an image... so that you don't have to get someone to like video it... because it can be chaotic and it can be a crisis.''} P2 felt that GenAI can depict actions realistically, because in \textit{``the first picture, she is rubbing sweaty hands on shirt, and she looks like super nervous.''}

\par
However, participants noted that some disability-related actions cannot be generated by GenAI. As P6 explained: \textit{``[I wanted to show] the person was not able to walk the stairs. [ChatGPT] didn't show that.''} Participants also observed some illogical or inconsistent actions across frames. For example, P3 pointed out that certain actions contradicted one another, stating: \textit{``One minute students were sitting and the next scene standing. Also if he is asking for someone to help write, why is he writing?''}

\textbf{Appearance Adjustment.} Participants P2, P5, P6, and P8 created GenAI prompts to refine the visual traits to align the characters with themselves. P2 wanted the character to have \texttt{``eyes open''} and to \texttt{``remove [NAME]'s glasses.''} P6 remade the image so that the character's \texttt{``eyes to look brown and not too skinny, medium.''} Participants also requested GenAI to customize clothing and styles to match the context. For example, P5's story took place in a theater, and they specified clothing details such as the character \texttt{``has a sparkly semi-formal dress''} and \texttt{``dark navy blue with short sleeves''}. Participants further wanted control over GenAI outputs to ensure the generated content accurately represented identity. In P6's creation, they asked ChatGPT to generate \texttt{``[NAME] should be a Black man, the rest of the people can be any color.''} P8 similarly requested the creation of an Asian male, stating, \texttt{``Remake the image, [NAME] wears glasses, Asian male with longer and straighter hair that somewhat falls into his face.''}
\par
\begin{figure}[!h]
    \centering
    \includegraphics[width=\linewidth]{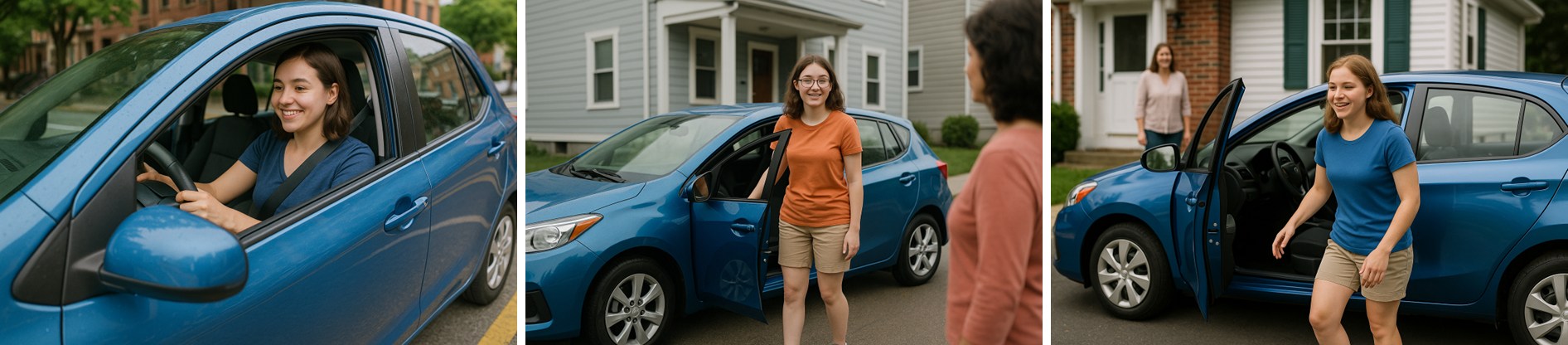}
    \caption{Left: ChatGPT created an scene of a PwD driving. Middle: In the next scene, ChatGPT changed the character's shirt color and added glasses. Right: P2 requested ChatGPT to restore the original shirt color and remove the glasses.}
    \label{fig:P2_1}
    \Description{Three images side by side. In the left image: a light-skinned adult woman with shoulder-length brown hair wearing a blue short-sleeved shirt. She is smiling as she drives a blue car. In the middle image: two light-skinned adult women, one on the left facing the camera, the other on the right facing away. On the left: the same woman as the previous image, but she is wearing an orange short-sleeved shirt, glasses, and tan shorts. She is smiling as she stands next to a blue car with the door open. On the right: the woman has shoulder-length brown hair  and is wearing a pink long-sleeved shirt as she looks at the other woman. In the right image: two light-skinned adult women, both facing the camera with a house behind them and a blue car with the driver door open between them. One is in the background on the left, the other is close to the camera on the right. On the left: the woman has shoulder-length brown hair and is wearing a pink short-sleeved shirt with blue pants as she looks at the other woman. On the right: the same woman as the previous images, she has shoulder-length brown hair wearing a blue short-sleeved shirt and tan shorts.}
\end{figure}

When creating the video, participants remade scenes because GenAI introduced inconsistencies in the character's appearance. These errors included variations in dress color, style, and the presence of accessories like glasses across different scenes, which disrupted narrative coherence. For instance, P2 noticed that ChatGPT had changed the character's shirt color and added glasses, prompting a request for revisions to maintain visual consistency (\autoref{fig:P2_1}). 

\par

In the interview, P5 and P6 expressed general satisfaction with a GPT-suggested appearance, noting, \textit{``one of the things that they made better, I didn't actually add, but it makes the person be in a suit.''} 
\par
However, some participants noted that GenAI may misrepresent PwDs or produce stories that feel inauthentic. P7 commented: \textit{``GenAI didn't really show people with disabilities well, like their bodies.''} P9 expressed a negative view of AI-generated personal stories: \textit{``It didn't sound genuine... it was artificially made.''} Similar to action inconsistency, participants also noted that the portrayal of PwDs lacked consistency across different GenAI images. P5 remarked, \textit{``One thing that I found challenging was when you try to type in a person, like the outfits or the person doesn't look the same across pictures.''}

\par

\textbf{Assistive Technology.} In video story creation, P1, P3, P4, P5, P7, and P9 prompted GenAI to depict assistive devices in relation to mobility, sensory, or learning needs. P1 instructed GenAI to create \texttt{``[NAME] walks slowly down the sidewalk using his walker.''} P4 requested an image of the character \texttt{``in an electric wheelchair waiting...''} Participants wanted GenAI to depict characters' experiences with assistive technologies in ways that reflected their own experiences. For example, P4 prompted GenAI with \texttt{``The driver should come out and say ``I would deploy the ramps on the train, but it would be too steep for you to get on. We would need a specific ramp.''''}
\par

\begin{figure}[!h]
    \centering
    \includegraphics[width=1\linewidth]{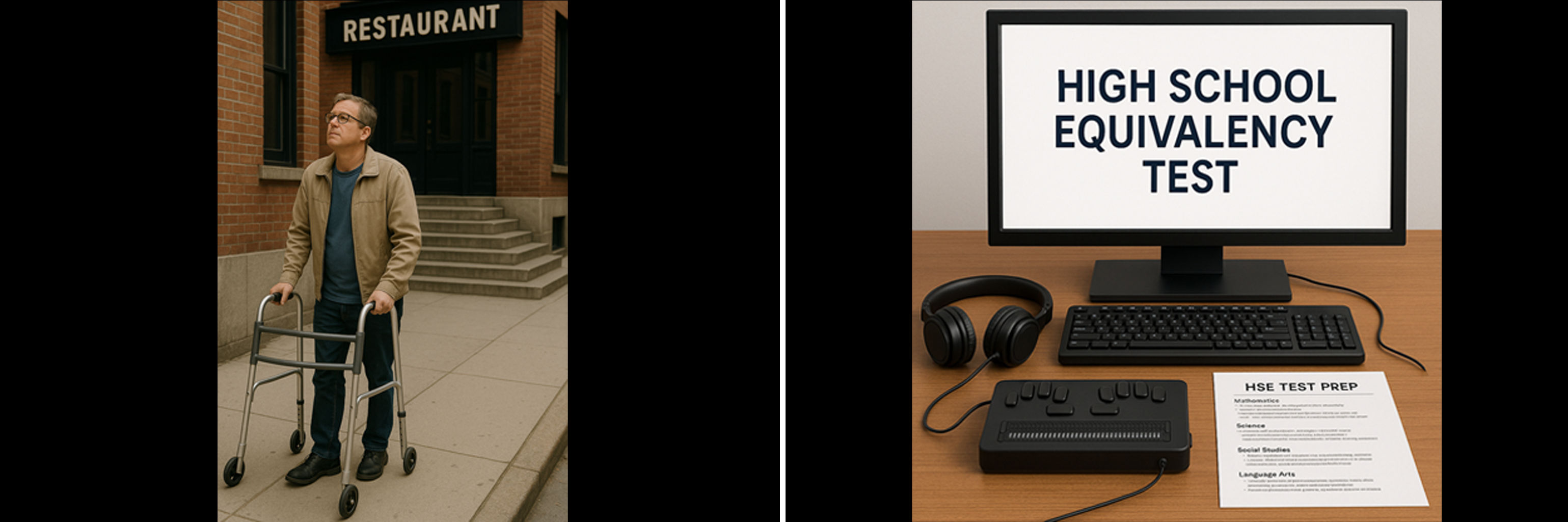}
    \caption{Left: P1's image with prompt \textit{``Create the same image but [NAME] should have a posterior-walker''.} Right: P9's image with the prompt \textit{``remake the image, it is not OGED. He uses a desktop not a laptop. The braille display should be plugged into a USB.''}}
    \label{fig:P1_P9}
    \Description{Two images, side by side. On the left: a light-skinned middle aged man with gray hair wearing blue jeans, a blue shirt, and a tan jacket uses an anterior walker as he walks towards the camera away from a Restaurant with stairs leading up to the door. On the right: a wooden desk with black headphones, a black braille screen-reader, a monitor and keyboard, and a piece of paper on top. The headphones are plugged into the screen-reader, which is plugged into the monitor. The monitor has the text: "High School Equivalency Test" displayed on it. The piece of paper has the text "HSE Test Prep" followed by the categories "Mathematics", "Science", "Social Studies", and "Language Arts".}
\end{figure}

However, certain specific assistive device descriptions were not accurately rendered by GenAI. For instance, in P1's creation, they requested a scene of a person using \texttt{``a posterior walker,''} but ChatGPT still produced only a standard walker (\autoref{fig:P1_P9} Left). ChatGPT also automatically associated certain assistive technologies with specific types of disabilities. For example, P3 prompted a story about \texttt{``someone named [NAME] with cerebral palsy who needs someone to help write in a college class,''} but ChatGPT depicted the character as a wheelchair user which was not expected by P3. In P9's video, the braille display was not plugged into the computer and the participant had to remake the images a few times (\autoref{fig:P1_P9} Right).
\par
In the interview, P1, P3, P5, and P9 particularly noted problems with generating correct assistive technologies using GenAI. P1 mentioned that \textit{``[ChatGPT] did not have a posterior walker.''} Participants emphasized that GenAI's ability to generate accurate assistive technology is important for ensuring that it is fair and free of bias. For example, P3 stated, \textit{``I feel [ChatGPT] automatically assumed someone was in a wheelchair... is not fair.''} P9 similarly noted, \textit{``It put [NAME] as wearing dark sunglasses... I see that as a bias.''}
\par
\textbf{Emotion Expression.} P1, P2, P3, P4, P5, P6, P8, and P9 included emotion-related descriptions in their prompts. Participants particularly prompted GenAI to depict nuanced emotions -- such as happiness, fear, frustration, and disappointment -- when facing disability barriers through facial expressions and body language. For example, in P2's creation, they asked ChatGPT to illustrate the character \texttt{``standing next to her blue car, legs trembling and sweaty hands wiping shorts,''} to convey fear and anxiety. In P5's video, they specified that the character should \texttt{``getting frustrated when they tell her the news (denied access to the theater), but tries to stay calm.''} Beyond the character's emotions, participants also used GenAI to visualize the emotions of people around them to illustrate their feelings of disability in a social context. In P6's story, they asked ChatGPT to \texttt{``an image of [NAME] speaking, and some lawmakers are keeping working, some do not care.''} P3 similarly prompted GenAI to create an image of the character \texttt{``struggling because people are just walking past and pretending not to see him.''}
\par
In the interview, P2, P3, P4, P5, and P8 specifically noted instances where emotions were accurately reflected in the GenAI images. For example, P3 remarked, \textit{``It did a good job showing him raising his hand, asking for help. It showed the woman coming over, so I felt it did really good on emotions here.''} Similarly, P6 observed, \textit{``It did a good job showing some of the people didn't like him.''}
\par
Participants also mentioned that the emotion expression felt ungenuine or incomplete. P9 noted that using GenAI to present PwDs' feelings \textit{``how it came out was artificial.''} P1 added that their feelings of frustration were not well depicted, suggesting that adding sounds like a harrumph could help convey the emotion more effectively.
\par

\begin{figure}[!h]
    \centering
    \includegraphics[width=1\linewidth]{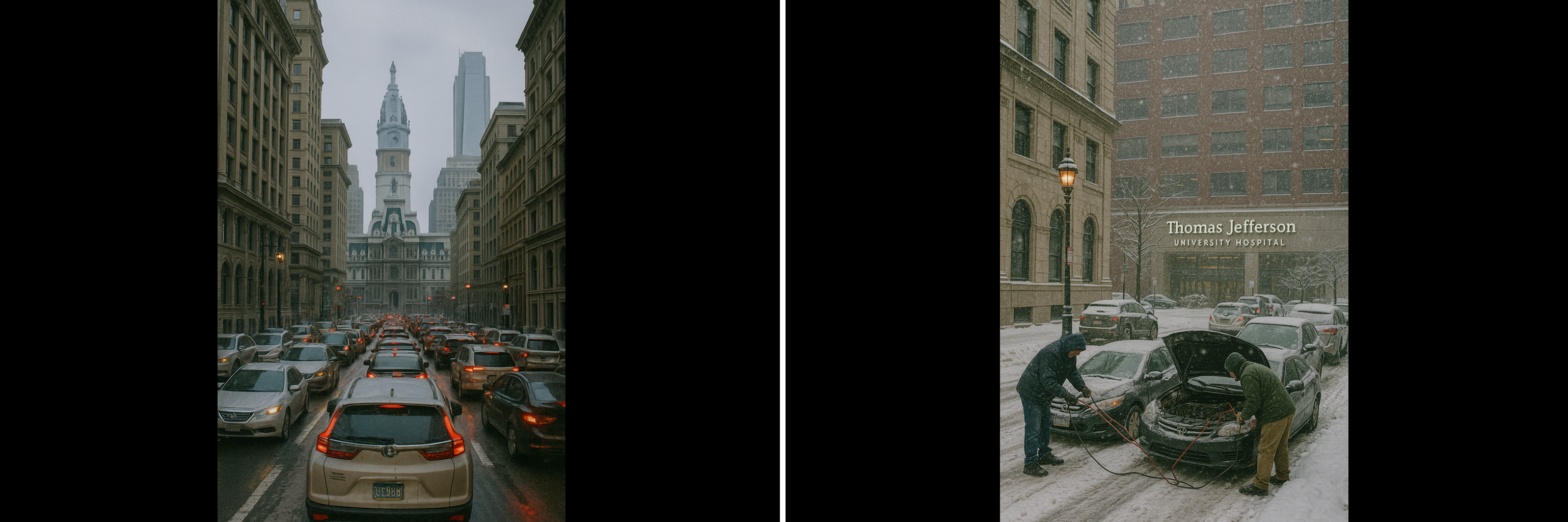}
    \caption{Left: Original frame created with P7's prompt, ``Create an image of Philadelphia the city and lots of cars crowded together.'' Right: Remade image with ``in Winter and people are trying to jumpstart their cars, and [at] Thomas Jefferson Hospital.''}
    \label{fig:P7_1}
    \Description{Two images side by side. On the left: many cars stuck in traffic on a wet city street. On the right: a city street covered in snow with Thomas Jefferson Hospital in the background. There are cars stopped in the background and in the foreground two people in winter coats are jumping a car as snow falls.}
\end{figure}

\textbf{Contextual Environment.} All participants prompted ChatGPT to create realistic accessibility-related environments. For example, in a story about the difficulties PwDs face walking on snowy days, P7 asked ChatGPT to create an image featuring Thomas Jefferson Hospital and to include the necessary cars for the video (\autoref{fig:P7_1}). P6 situated an experience at the Massachusetts State House to illustrate a disability advocacy scenario. Participants also used GenAI to illustrate places with poor accessibility. P4 prompted GenAI to create an image of a Boston Green Line station that lacked a ramp for their wheelchair. P3 prompted the creation of a classroom where their character does not receive needed support. PwDs further used GenAI to depict how the environment impacts their mental health. For example, in P8's story about sensory stress, they prompted: \texttt{``Create an image of a street with bright long light poles all lit, a lot of people walking... very noisy.''}
\par

\begin{figure}[!h]
    \centering
    \includegraphics[width=1\linewidth]{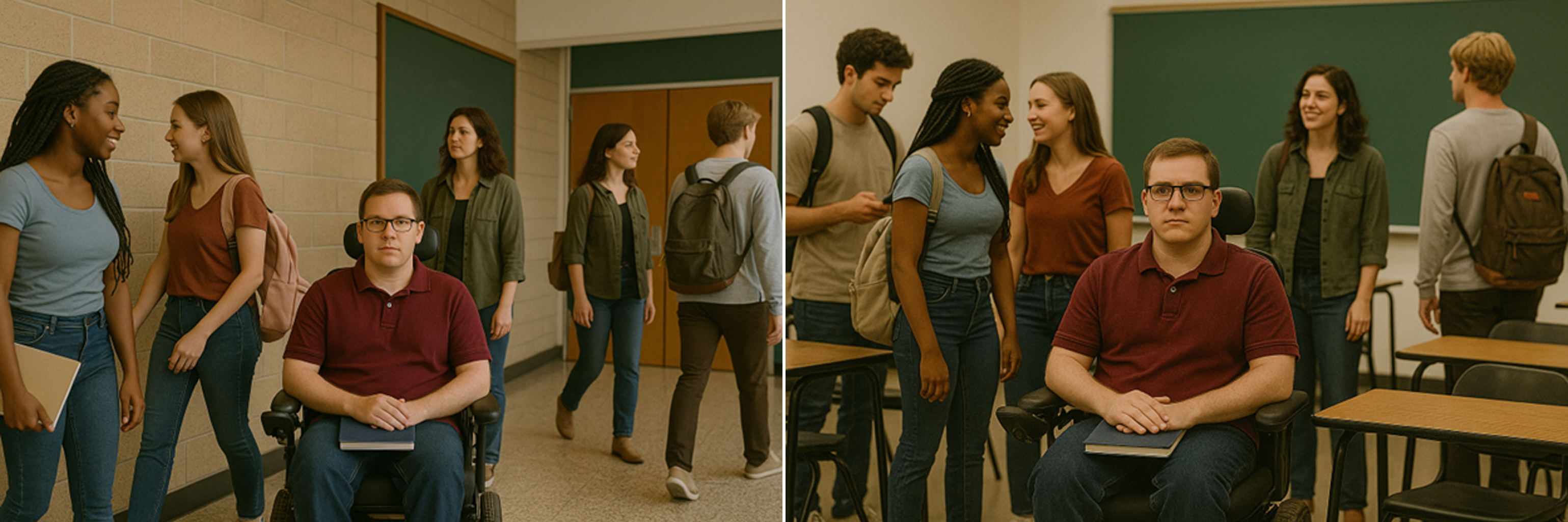}
    \caption{Left: The original scene generated by ChatGPT, where [NAME] (P3) is not shown in the classroom as indicated in the previous scene. Right: The revised scene after P3 specified, \texttt{``remake the image but [NAME] should be in the classroom.''}}
    \label{fig:P3_1}
    \Description{Two images side by side. On the left: a light-skinned adult man with short brown hair wearing blue jeans and a red collared shirt using an electric wheelchair faces the camera outside in a school hallway. Students walk behind and next to him as he has his hands crossed over a blue notebook on his lap. On the right: the same man as the previous image faces the camera inside of a classroom. Other students stand and talk behind him.}
\end{figure}

However, participants noted difficulties when GenAI made unexpected changes to the story environment, such as altering the location, weather, or visual style. For example, ChatGPT generated an image of the character sitting in a hallway in P3's story, even though the story took place in a classroom (\autoref{fig:P3_1}). In P7's story, ChatGPT failed to depict the scene of a snowy day and produced an image that did not show the outside of the hospital with snow.

In the interviews, participants appreciated the flexibility GenAI provides in selecting environments and exploring how those environments feel. P7 noted, \textit{``I just thought it would be interesting to see how [the hospital in Philadelphia] looked.''} P5 similarly remarked that \textit{``you can pick any... backdrop scene that you want.''} Participants emphasized the importance of using GenAI to depict real locations in order to raise administrators' awareness of inaccessibility. For example, P4 noted that \textit{``if the ramp is too steep... cities or towns need to fix it so that people with wheelchairs can ride through.''} P5 added, \textit{``There's so many theaters built a long time ago and don't have elevators.''}

\subsubsection{GenAI Media Quality and Efficacy}
During the study, P2, P5, and P9 created prompts to improve image settings and quality. P2 requested GenAI to make the image in a \texttt{``more realistic style.''} P9 asked ChatGPT to \texttt{``remake the image so there are two separate, in one he is not holding the instructors..., in the other he is in mathematical quicksand.''}

\par
\begin{figure}[!h]
    \centering
    \includegraphics[width=1\linewidth]{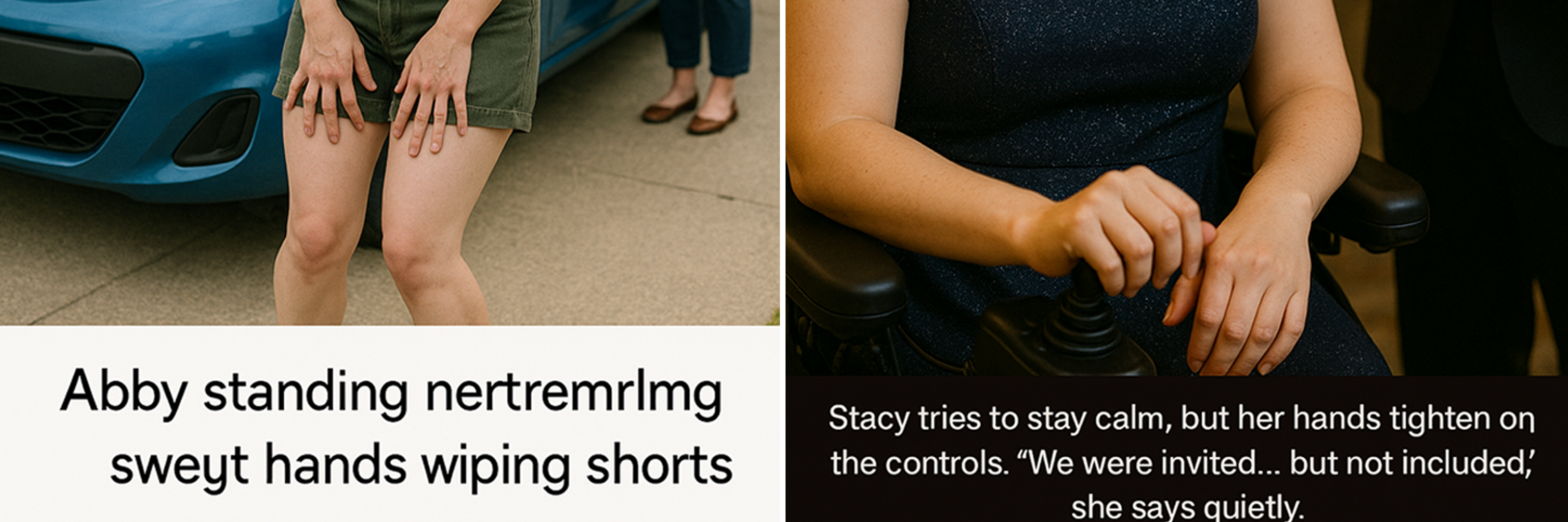}
    \caption{ChatGPT-generated scenes include unexpected text with misspellings.}
    \label{fig:P2_P5_1}
    \Description{Two images side by side. On the left: a cropped image with two pairs of legs in front of a blue car. One is in the foreground, one is in the background. The legs in the foreground are wearing shorts and have hands being wiped on them. The legs in the back have blue pants and brown shoes on. Below the image is some text: "Abby standing nertremrlmg sweyt hands wiping shorts". On the right: a cropped image of a wheelchair user in a dark blue sparkly blue dress, below their shoulders to their knees are visible. Below the image is some text: "Stacy tries to stay calm, but her hands tighten on the controls. 'We were invited... but not included, she says quietly."}
\end{figure}
\begin{figure}[!h]
    \centering
    \includegraphics[width=\linewidth]{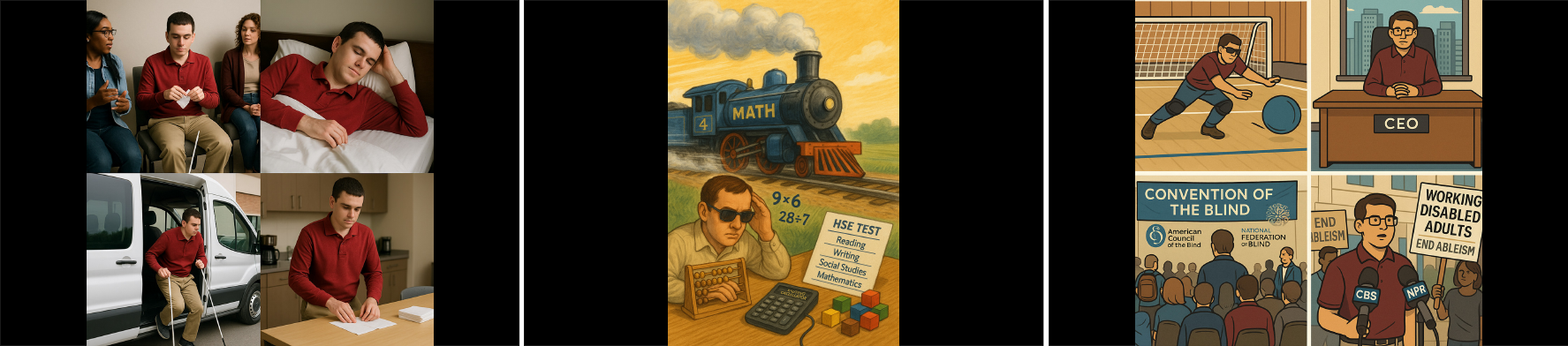}
    \caption{Three frames from P9's story, shifting from a realistic four-frame image to an artistic style, and then to four artistic frames.}
    \label{fig:P9_1}
    \Description{Three images side by side. On the left: a four frame frame picture, in the top-left frame a light-skinned adult man with short black hair wearing a red collared shirt and tan pants sits in a gray chair. A white cane rests on his leg as he tears up a piece of paper. Two women are in the image, one one either side of him, both also sitting in gray chairs. In the top-right frame the same man lies in bed under white sheets on a white pillow, still wearing a red collared-shirt. In the bottom-left frame the same man gets out of the back of a white van, using a white cane in each hand. In the bottom-right frame, the same man wearing the same outfit folds napkins on a wooden table, a kitchen sink and a coffee machine on the counter behind him. In the middle: a drawing-stlye image of a light-skinned man with short brown hair, wearing a white collared-shirt and black sunglasses sitting at a table. On the table is an abacus, a calculator, a pile of colored wooden blocks, and a piece of paper. The calculator displays the text: "calculator". The paper has the text: "HSE TEST" and the sections "Reading", "Writing", "Social Studies", and "Mathematics". The equations nine times six and twenty-eight divided by 7 float above the man's head. Behind him is train track with a blue strain on it, the number four and the word "MATH" are written on the side in yellow. On the right: another four-frame picture, but in drawing-style. In the top-left frame a light-skinned adult man wearing a red short-sleeved shirt, blue pants, kneepads, and black sunglasses plays goal ball on a wooden court with a goal behind him. In the top-right frame the same man now wearing a red collared-shirt and black glasses sits at a desk labelled "CEO", a window showing a city skyline behind him. In the bottom-left frame, a crowd of people face away from the camera, looking at a sign with the text: "Convention of the blind", "American Council of the bind", and "National federation of blind". In the bottom-right frame, the same man shown in the previous frames, wearing a red collared-shirt and black glasses, speaks into two microphones. One is labelled "CBS" and the other "NPR". Behind him two people stand, each holding signs. One has the text: "End Ableism", the other: "Working Disabled Adults End Ableism".}
\end{figure}

However, ChatGPT sometimes added unwanted, misspelled text to the images (\autoref{fig:P2_P5_1}). In response, participants instructed ChatGPT to remove the text. The GenAI image style also occasionally changed without participants' instructions. In P9's story, the frames shifted from a realistic four-frame image to an artistic style, and then to four artistic frames (\autoref{fig:P9_1}).

In the interview, all participants commented on GenAI media quality and efficacy. Some praised the quality and efficiency of GenAI-generated scripts, images, and voiceovers. For example, P2 stated, \textit{``I was very impressed with how good the images were.''} P1 highlighted the efficiency of GenAI media, noting, \textit{``I don't have to like waste time on using my voice as a narrator.''} 

Participants also mentioned that GenAI provided helpful guidance in creating the story and making sense of its structure. P2 added: \textit{``It helps start... sometimes it's just hard to start a project or story.''} Similarly, P4 shared: \textit{``[GenAI] is like a tutor for anybody that wants to tell their story. Even with people that have disabilities, like myself, I have ASD and I don't wanna make stories that don't make sense, so I find this very useful.''} P9 noted that GenAI did a good job explaining the content to them. 
\par
Participants also noted challenges related to the need to specify detailed instructions for GenAI to create images. P1 remarked that \textit{``[ChatGPT] don't necessarily get the minor details exactly right unless you put them in.''} While our task utilized only static images and monologue voiceovers, participants expressed a desire for GenAI to generate more diverse forms of media. For example, P6 remarked that when using GenAI videos, \textit{``I wish they actually did like show I am talking. I wish like the [image] was moving.''} More importantly, participants commented on the tension between receiving support from GenAI and maintaining ownership of their voice and creativity. P3 emphasized that GenAI should not take PwDs' voice, and P9 noted, \textit{``it wasn't really giving me the creativity that I wanted... You're asking the computer to do the thinking for you.''}

\subsection{RQ3: Sharing Purposes of GenAI-Created Stories}

For RQ3, the interview questions asked how participants felt about sharing these stories on social media and with whom they would feel comfortable sharing them. As previously discussed, participants noted that GenAI-produced disability story videos could lack genuine emotion and sometimes misrepresent PwDs' bodies or assistive technologies. They suggested improving viewer engagement by adding animation (P6), music (P8), or stylistic elements such as drama (P9). However, most participants still indicated they would share the videos with the general public, family and friends, and property owners. Participants emphasized that these issues did not overshadow the benefits of raising public awareness, sharing communal feelings, and initiating conversations with property administrators.

\subsubsection{General Public on Social Media}
All participants commented on how they felt about sharing the videos with the general public on social media. P1, P2, P3, P4, P5, P6, and P7 expressed positive opinions about using GenAI to raise disability awareness. For example, P7 mentioned that the GenAI video could contribute to their social media presence, noting, \textit{``[GenAI video] would probably help my YouTube channel get going.''}
\par
P8 and P9 felt uncomfortable sharing the videos publicly on social media because the GenAI content might not be attractive enough to draw public attention. P8 chose not to share because they felt the GenAI video quality was not good enough. P9 felt that sharing the GenAI video on social media \textit{``is not changing anything about disability... If [the story] is not being seen on CBS or ABC or Fox News or NBC, I think it is pointless.''}

\subsubsection{Family and Friends}
Five participants, P2, P3, P4, P5, and P8, indicated that GenAI videos could be shared with family and friends. Participants noted that sharing the GenAI video is a novel way of explaining what is happening and highlighting accessibility needs in the community (P5), and can help PwDs better understand who they are (P8). Participants also expressed interest in using GenAI-created videos to foster community bonding. For example, P2 stated, \textit{``I feel like there [are] some trying to go for [driving] permits and licenses, and [the GenAI video] would show them that they are able to do it even when we are nervous.''}

\subsubsection{Property Administrator}
Three participants, P1, P5, and P9, hoped their videos would be viewed by property administrators of government buildings, businesses, schools, or other public facilities because they saw the GenAI story video as an early alert that could prompt greater attention to accessibility during construction design. For example, P1 stated, \textit{``Business owners. [GenAI videos] maybe give them a reason why businesses need to be made accessible for people with disabilities.''} Participants also hoped that GenAI-facilitated videos could act as conversation starters to confront accessibility challenges. P5 mentioned that \textit{``they need to know there is still disability issues they need to get worked on.''} Although P9 would not share the video publicly, he noted that if administrators viewed it, it could start a conversation and help challenge stereotypes.

\section{Discussion}
\begin{figure*}[!h]
    \centering
    \includegraphics[width=0.9\linewidth]{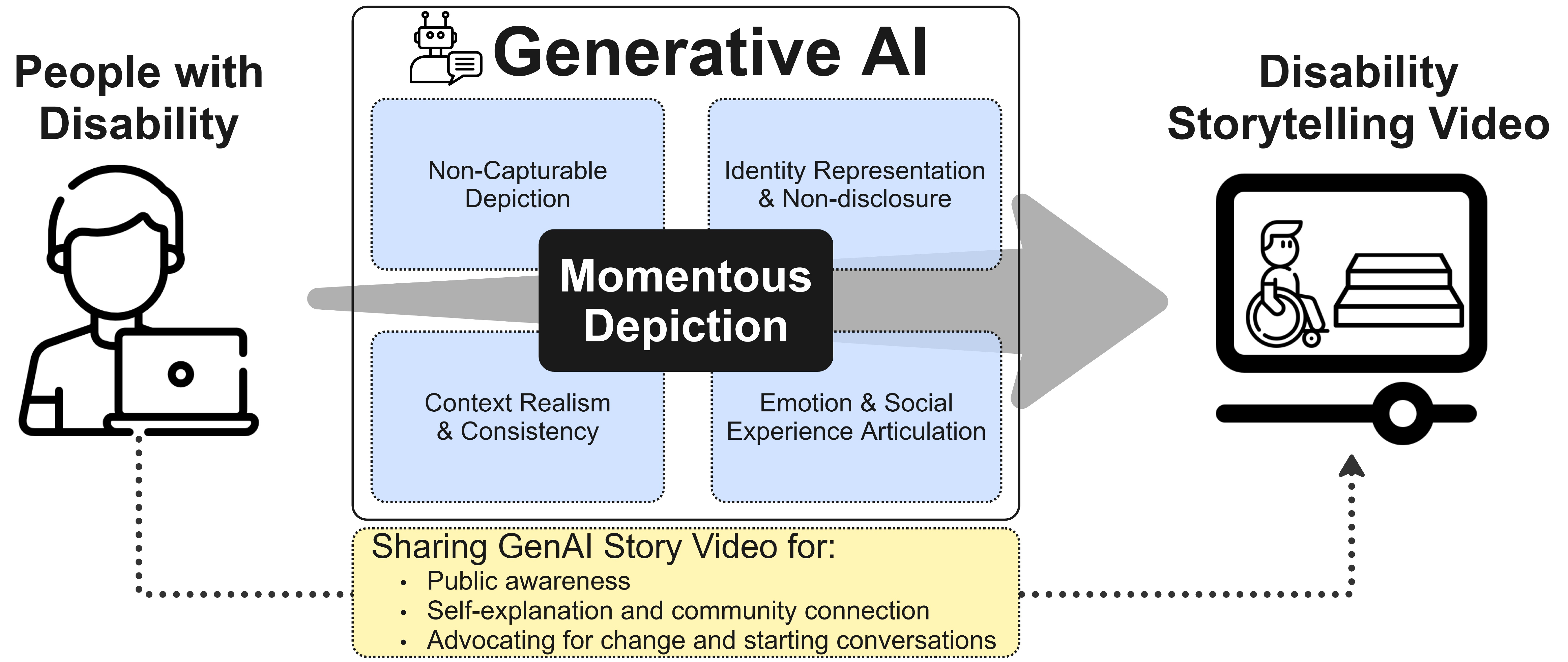}
    \caption{A conceptual framework of Momentous Depiction that highlights the GenAI affordances needed to support disability storytelling.}

    \label{fig:framework}
    \Description{A diagram with a person on a computer labelled "People with Disability" on the left, a box labelled "Generative AI" in the center, and a screen labelled "Disability Storytelling Video" on the right. An arrow points from Disability Storytelling Video to People with Disability, labelled "Disability Story Sharing" with the sub-text "Self-expression and connection" and "Public Awareness". The box in the center contains five text elements, "Non-capturable Depiction", "Identity Non-disclosure", "Context Realism and Consistency", emotion \& social experience articulation", and above them all "Momentous Depiction".}
\end{figure*}

Based on our findings, we conceptualize GenAI as a design material for \textit{momentous depiction} (\autoref{fig:framework}) -- a tool for visualizing critical moments that convey the insights and meanings of disability~\cite{lambert_digital_2013}. In HCI, AI is regarded as a critical design material that introduces new affordances but also imposes constraints~\cite{yildirim_how_2022}. Within this framework, GenAI's momentous depiction must provide four critical affordances: \textit{non-capturable depiction}, \textit{identity representation and undisclosure}, \textit{context realism and consistency}, and \textit{emotion \& social experience articulation}. By depicting such moments, PwDs employ GenAI for public awareness, self-expression, and advocacy. This framework highlights not only how PwDs without rich media or technology experience can integrate GenAI content into their narratives, but also the GenAI challenges that remain to be resolved.

\subsection{Affordance 1: Non-Capturable Depiction}
\textbf{Participants noted that GenAI affords the depiction of moments that are difficult to capture in real life.} During our study, they were impressed by GenAI's efficiency in creating realistic images of contexts, characters, and emotions. They used GenAI to visualize real-world experiences that are difficult to record, such as being denied access or experiencing intense stress. GenAI was also used to depict wishful futures or societal support~\cite{angelini_criptopias_2023}. Such depiction is achieved through carefully configured actions and contextual environments. Depicting such moments was seen as a way to exchange disability experiences, foster community, and connect with other PwDs. As part of their roles, participants noted that visualizing otherwise non-capturable moments can raise awareness and initiate conversations.
\par
However, these depictions were also challenged by imperfections in GenAI~\cite{weisz_design_2024}, including unwanted content (e.g., text added without prompting) and illogical behaviors (e.g., a character writing while asking for writing support). Participants suggested that depicting non-capturable moments required greater flexibility in media formats, such as choosing different visual styles or using split images.
\par

Sharing disability-related stories plays a critical role in communicating with non-disabled audiences~\cite{seo_understanding_2018, cocq_self-representations_2020}. HCI research has incorporated tools such as live cams and video recordings to support story recording~\cite{rodger_journeycam_2019, hibbard_vlogging_2011, jiang_its_2024}. Our findings suggest that leveraging GenAI to reflect on and visualize specific moments in the lives of PwDs can expand the range of experiences that can be captured and thereby enhance disability communication. Such affordance reflects Dourish's notion of situated action, where the meaning of disability stories emerges through enriching specific contexts and interactions with GenAI~\cite{dourish_where_2004}.

\subsection{Affordance 2: Identity Representation and Non-disclosure}
\textbf{Participants co-created new disability identities based on real or envisioned experiences without needing to disclose their actual identities}. While prior HCI studies note that PwDs selectively choose what to disclose based on context and communication needs~\cite{porter_filtered_2017, zhang_its_2022, gualano_i_2024, rauchberg_shadowbanned_2022}, GenAI enabled a new media format in which identity is fabricated while the disability experience remains authentic. Our study identifies five selective configurations for presenting disability identities with GenAI: the actions related to disability experiences, the character's appearance and racial background, the momentous emotions, the assistive technology depicted, and the context of disability challenges. While prior work has noted biased representations in GenAI tools~\cite{mack_they_2024, glazko_autoethnographic_2023}, our findings indicate that these tools remain valuable for developing new forms of disability disclosure in public spaces~\cite{halperin_probing_2023, hofmann_living_2020, boyd_exploring_2024}. Therefore, participants emphasized that PwDs should maintain autonomy in deciding the story ideas through which they share their voices. During GenAI-supported identity construction, GenAI should facilitate reflection on disability identities and offer guidance rather than constructing the entire story.

\par
However, participants experienced breakdowns when co-creating disability identities with GenAI. The storytelling experience was hampered when imagined characters or assistive technologies could not be rendered. ChatGPT misrepresented individuals with specific disabilities by failing to depict certain assistive devices (e.g., inability to generate a posterior walker) or showing stereotypes (e.g., defaulting to wheelchair use for PwDs with cerebral palsy). These issues contributed to participants' perceptions of bias in GenAI tools~\cite{mack_they_2024}.
\par
Prior studies have noted that GenAI often produces biased and disrespectful portrayals of PwDs~\cite{mack_they_2024, glazko_autoethnographic_2023}. Our findings further confirm that such limitations may discourage some PwDs from using GenAI for storytelling. This challenge is especially critical for individuals with non-apparent disabilities~\cite{bitman_which_2023}. GenAI tools must therefore be designed with careful consideration of how diverse disabilities are portrayed in GenAI outputs. 

\subsection{Affordance 3: Context Authenticity and Consistency}
\textbf{In momentous depiction, participants emphasized using GenAI to localize scenes and render authenticity and realism in storytelling.} Though GenAI content is fake by nature, participants still prompted GenAI visuals to present stories in real-world locations, such as the Massachusetts State House, Thomas Jefferson Hospital, and the Green Line in Boston. Such realism aims to depict real-world scenarios and convey targeted messages to relevant stakeholders.

\par
This contrasts with prior research showing that LLMs frequently default to ``inspiration'' or ``disability porn''~\cite{gadiraju_i_2023, mack_they_2024, gogolushko_ai_2022}. Our study suggests that it is critical to reconsider how GenAI content can form PwDs' sense of authenticity~\cite{nevsky_each_2025}. Participants noted that one reason for not sharing GenAI content was a feeling of inauthenticity. When communicating disability issues~\cite{chheda-kothary_engaging_2024} and raising awareness~\cite{iniesto_designing_2020, boyd_exploring_2024}, the affordance of grounding stories in real-world settings can help align AI-generated narratives with PwDs' personal experiences~\cite{bennett_painting_2024}.
\par
However, GenAI's inconsistencies and biases threaten the authenticity of storytelling. The appearance of AI-generated characters -- such as their clothing and accessories -- shifted across scenes, even when the same storyline and character name were specified. GenAI visuals also presented illogical actions or emotions. Environmental settings and background characters changed unnaturally between scenes. These inconsistencies made it difficult to assemble coherent narratives. 

\par
HCI research on disability storytelling has emphasized media richness~\cite{stangl_transcribing_2015, yoo_remembering_2024, jiang_its_2024} and interactivity~\cite{rodger_journeycam_2019, iniesto_designing_2020}. Research on GenAI accessibility has primarily focused on interface design and mitigating discriminatory outputs~\cite{mack_they_2024, glazko_autoethnographic_2023, gadiraju_i_2023}. Adding to these considerations, in GenAI-supported storytelling, our findings highlight authenticity and consistency as emerging criteria for enabling momentous depictions. GenAI's lack of global awareness is a known challenge in other creative domains~\cite{bennett_painting_2024, lee_altcanvas_2024}. Future research should explore how improving AI's global awareness could help resolve consistency issues in storytelling.

\subsection{Affordance 4: Emotion \& Social Experience Articulation}
In momentous depiction, participants used GenAI to illustrate emotions and social experiences related to their lived disability experiences. GenAI's ability to generate facial expressions, body language, and social context was particularly used by participants to express feelings about disability and convey their messages. We also note that PwDs particularly depicted social experiences by showing the emotions and reactions of people in the surrounding environment -- an element that is both unique and critical in disability storytelling, as one goal of sharing the GenAI videos was to promote social acceptance. Depicting emotions and social experiences through GenAI can raise awareness among family members, friends, and the general public, helping others better understand their lived experiences.
\par

This finding highlights that the affordance of accurately presenting emotion is essential for GenAI-generated images in disability reflection and communication. Emotional expression is central to disability storytelling~\cite{yoo_understanding_2021, yoo_remembering_2024}, and PwDs may engage with AI agents to process emotions and navigate the harms of ableism~\cite{mcnally_disability_2024}. Our study extends these findings by demonstrating that GenAI's capacity for emotional presentation plays a critical role in storytelling. Consistent with research on emotional disclosure in storytelling on social media~\cite{niu_please_2024, seo_understanding_2018}, our findings suggest that PwDs expect GenAI characters and their social environments to accurately convey intended disability experiences.

\subsection{Design Implications}
Based on the four key affordances of GenAI, we propose three primary directions for future HCI design and research on integrating GenAI into systems that support video-based disability storytelling.

\subsubsection{Incorporating GenAI Affordances for Story Completion}

In achieving momentous depictions with GenAI, our findings suggest that designs should prioritize how GenAI enhances PwDs' autonomy in crafting stories of their own lived or envisioned experiences while GenAI fills in supporting details, rather than producing full storylines or automating the entire storytelling process. As GenAI should support PwDs' personal expression~\cite{glazko_autoethnographic_2023, bennett_painting_2024, mcnally_disability_2024, choi_exploring_2025, lee_altcanvas_2024, s_guedes_artistic_2024}, our results suggest the opportunity to support writing and generate non-capturable, localized, and emotional visuals with GenAI. Prior research has often conceptualized GenAI as co-authors or collaborators in creative practices~\cite{bennett_painting_2024, hu_designing_2025, yildirim_how_2022}. In the storytelling context, the role of GenAI should not be viewed as that of a ``story writer'' but rather as a ``story completer.'' The notion of a ``story completer'' underscores the contributions of PwDs through their contexts, lived experiences, and narrative intentions, while GenAI provides supplementary details that enhance authenticity and localization. This perspective extends digital storytelling theory by reinforcing the agency of human storytellers over AI systems~\cite{xiao_sustaining_2025}.
\par
In using GenAI for story completion, another necessary feature is the ability to adjust momentous elements so they align with intended narratives. We argue that GenAI for PwD storytelling must depict not only what people do, but also what they \textit{cannot} do (e.g., being unable to walk up stairs). Such capabilities require GenAI features that support revising story details and refining prompts to generate more accurate images. Future GenAI systems should also facilitate prompt construction through structured guidance and illustrative examples that foster users' imagination~\cite{angelini_criptopias_2023, chopra_storybox_2022, fina_storytelling_2016}. For instance, as PwDs create playful videos to depict daily life and challenge stereotypes~\cite{duval_chasing_2021}, GenAI should likewise support the representation of positive experiences, including play and leisure.
\subsubsection{Incorporating GenAI Affordances as a New Identity-Concealing Media Format}
The incorporation of GenAI introduces a novel media format for reflection and imagination, enhancing PwDs' ability and willingness to express. Within technologies that support moment capturing~\cite{yoo_remembering_2024, hibbard_vlogging_2011}, reminiscing~\cite{yoo_understanding_2021}, and disability disclosure~\cite{iniesto_designing_2020} -- which have traditionally relied on camera recording or text drafting -- GenAI-generated stories can serve as a new design material and media type that complement these practices. This format can potentially help address challenges related to visibility~\cite{choi_its_2022, sannon_disability_2023}, exposure to harassment, and difficulties in adapting content strategies for public dissemination~\cite{sannon_disability_2023, miller_my_2017}. Designers can incorporate GenAI-generated media as an alternative mode of self-presentation and identity non-disclosure~\cite{zhang_its_2022, porter_filtered_2017, gualano_i_2024}: GenAI functions as a story visualizer that scaffolds PwDs' expression of experiences~\cite{hu_designing_2025} while enabling them to conceal personal profiles and information.
\par
GenAI in prior work has primarily focused on general media creation or artistic expression for PwDs~\cite{choi_exploring_2025, lee_altcanvas_2024, s_guedes_artistic_2024}. Such designs should also consider how GenAI can foster a sense of authenticity and realism. GenAI-supported storytelling tools could incorporate video styles from professional creators~\cite{bartolome_literature_2023} to more effectively present characters, convey contextual emotions, and depict realistic environments. GenAI's limitations in accurately profiling PwD characters and presenting correct assistive technologies must be addressed to enhance authentic disability experiences. 
\par
Prior research has shown that viewers often perceive GenAI content as lower quality~\cite{rae_effects_2024}, prompting social media platforms to restrict AI use~\cite{lloyd_ai_2025} and introduce labels to flag such content~\cite{gamage_labeling_2025}. However, our study suggests that supporting PwDs in leveraging GenAI for authentic self-expression is promising. Platforms should reconsider how the visibility of PwDs' GenAI content is managed so that their GenAI-facilitated narratives are not unfairly suppressed.

\subsubsection{Designing GenAI Affordances as an Imperfection Corrector}
Designing for imperfection~\cite{weisz_design_2024} is a critical principle in developing human-centered GenAI applications. In the context of GenAI-supported momentous depiction, GenAI tools should be designed to guide PwDs in recognizing and correcting biased representations, incorrect logics, and inconsistencies. Addressing issues of disability representation~\cite{mack_they_2024} requires inclusive AI design practices, wherein PwD communities co-develop standards and guidelines for disability identity~\cite{costanza-chock_design_nodate}. Such collaboration can yield practical methods that help PwDs craft effective prompts. To advance inclusive storytelling, we call on AI researchers to collaborate with PwD communities to establish standards for how different disabilities should be profiled by GenAI.


\par

To address the need for consistency in momentous depiction, GenAI tools should provide features that allow users to lock or persist character traits, clothing, and environmental settings across scenes. Prior studies have highlighted challenges with GenAI content failing to align with creative intent and nuanced experiences~\cite{bennett_painting_2024, mcnally_disability_2024}. Extending these concerns, we recommend that new GenAI designs to ensure coherence of story materials across multiple outputs. Potential solutions include reusable AI character templates, enhanced visual memory for scene continuity, and prompt history anchoring to minimize inconsistencies.

\section{Conclusion}
This research provides insights into how GenAI can support PwDs in creating storytelling videos about disability. We found that participants held positive attitudes toward using GenAI for storytelling, as it enhanced media-creation efficacy and helped depict non-capturable moments. We propose the Momentous Depiction framework to highlight four affordances that support GenAI-enabled storytelling while also revealing limitations in GenAI: depicting non-capturable moments while occasionally having illogical or unshowable actions; presenting identities in an undisclosed manner while sometimes misrendering disabilities and assistive technologies; visualizing disability story contexts while still exhibiting inconsistencies and biases; and depicting emotions and social experiences, though the emotional tone may feel ungenuine or incomplete.

\section{Limitations and Future Work}
Our study has several limitations. First, while participants shared similar experiences in disability advocacy and did not represent the full diversity of PwDs, future work should examine how GenAI storytelling practices vary across mobility, speech, and hearing disabilities. The Momentous Depiction framework can guide such work by exploring GenAI's capabilities and limitations in depicting different non-capturable disability moments and how various groups perceive contextual realism. 
\par
Second, our study focused on novice video creators; experienced disabled creators may engage with GenAI differently due to their familiarity with editing tools, platform norms, and audience strategies. Our framework also raises questions about how professional creators present moments, identities, contexts, and emotions compared to novice creators, and how GenAI can support such performed practices.
\par
Lastly, we examined only ChatGPT, DALL·E, and ElevenLabs. As new GenAI tools emerge, future research should explore how new multimodal affordances enhance or constrain disability storytelling. Advanced AI video generation models such as Sora~2 and Veo~3.1 were released after our study. With the rapid evolution of GenAI technologies, it remains essential for HCI and accessibility researchers to continue examining their roles. Researchers and platforms should consider how malicious actors might misuse these tools to spread stereotypes against PwDs.

\par

\section{Acknowledgments}
The authors used LLM only for grammar checking and copyediting, without contributing ideas or generating core findings. The images presented in the Results section were created with DALL-E based on participant-provided story prompts during the study.

\bibliographystyle{ACM-Reference-Format}
\bibliography{references}

\end{document}